\documentclass[10pt, conference]{IEEEtran}
\usepackage{tikz}
\usepackage{amsmath,amsfonts}

\usepackage{graphicx}
\usepackage{subcaption}
\usepackage{colortbl}
\usepackage{hyperref}
\usepackage{listings}
\usepackage{multirow}
\usepackage{pifont}
\usepackage{textcomp}
\usepackage{xcolor}
\usepackage{xspace}
\usepackage[font=small,labelfont=bf]{caption}
\usepackage{booktabs}
\usepackage{algorithm}
\usepackage{algpseudocode}
\usepackage{times}

\definecolor{commentcolor}{RGB}{102, 153, 0}

\newcommand{\nxt}[1]{\textcolor{black}{#1}}
\newcommand{\dtt}[1]{\textcolor{black}{#1}}     % NXT ++ Choose a dark green for better visualization

\newcommand{\remark}[1]{\textcolor{black}{#1}}

\newcommand{\sysname}{DPS\xspace}   %% Dual-Mode Precision LLM Serving

\newcommand{\woff}{\textsc{Weight-Offload}\xspace}
\newcommand{\kvquant}{\textsc{KV-Quant}\xspace}
\newcommand{\kvoff}{\textsc{KV-Offload}\xspace}

\newcommand{\FullMode}{\textbf{full mode}\xspace}   
\newcommand{\PerfMode}{\textbf{fast mode}\xspace}

\IEEEoverridecommandlockouts
\usepackage{orcidlink}  % For Orcid

\begin{document}

 \title{\sysname{}: Dual-Mode Precision LLM Serving with Semi-Unified Memory}
%\title{\sysname{}: Dual-Precision LLM Serving with Semi-Unified Memory}
%{\footnotesize \textsuperscript{*}Note: Sub-titles are not captured for https://ieeexplore.ieee.org  and should not be used}
\author{Xuan~Truong~Nguyen\orcidlink{0000-0002-7527-6971}\textsuperscript{*},~\IEEEmembership{Member,~IEEE}, 
Tien Son Pham\textsuperscript{*}, Tuan Duc Chu, 
Wookeun Jung, and Thanh Tuan Dao\orcidlink{0000-0002-9897-2769}
\thanks{This work was supported by Moreh and Van-Lang Institute of Semiconductor Technology (VIST). (\textit{Corresponding author: Thanh Tuan Dao}.)

Xuan Truong Nguyen is with the Department of Next Generation Semiconductor Convergence and Open Sharing System (COSS), Seoul National University, Seoul 08826, South Korea. He is also with Van-Lang Institute of Semiconductor Technology (VIST), Hanoi, Vietnam. (E-mail: truongnx@snu.ac.kr).
%, Hanoi, Vietnam (E-mail: truongnx@snu.ac.kr).

Tien Son Pham and Tuan Duc Chu were with the Efficient Computation Research Group, Moreh Vietnam. (Email: \{phamtienson02, chutuanduc0505\}@gmail.com). Wookeun Jung and Thanh Tuan Dao are with Moreh. (E-mail: \{wookeun.jung, tuan.dao\}@moreh.io)
} 
\thanks{* Equal contribution.}
}
% \and
% \and
% \and
% \and
% \and

\maketitle

\begin{abstract}
Existing LLM serving systems virtualize and optimize KV-cache memory, but treat model-weight memory as fixed throughout execution. Recent work on multi-precision model representations challenges this design by allowing a single stored model to support both full-accuracy and lower-precision execution, making the effective weight footprint runtime-dependent. This creates an opportunity under bursty workloads, where temporary spikes in KV-cache demand often determine throughput and SLO compliance. We present DPS, a dual-precision LLM serving system that turns weight memory into an elastic resource: under normal load, DPS serves the full-accuracy model; under KV pressure, it switches to a nested, lower-precision variant and repurposes unused weight memory for KV cache blocks. DPS is built on Semi-Unified Memory (SUM), which partitions the weight region into a persistent lower-precision sub-region and a shared region that alternates between residual weight tensors and KV-cache blocks, preserving compatibility with paged KV-cache management. 
We implement DPS on top of vLLM and evaluate it across both dense and MoE models and various production workload traces. Our results show that \sysname improves sustained throughput by $2.1$--$3.3\times$ and effective pass@1 by up to $+41$\,pp over Static FP16, while preserving FP16-class accuracy.
\end{abstract}

\section{Introduction}\label{sec:introduction}
Transformer-based large language models (LLMs) such as GPT~\cite{gpt3}, Llama~\cite{llama}, Qwen~\cite{qwen1.5}, and DeepSeek~\cite{deepseekai2024_deepseekv2_strongeconomicalefficient, deepseekai2025_deepseekv3_technicalreport} have been emerging as foundation components in modern AI services, including virtual assistants, chatbots, text, image, and code generation~\cite{chatgpt, copilot}.
Their strong performance across various domains has attracted significant attention and driven significant user demand. %, as reported by the industry~\cite{}
\dtt{
%To meet system-level objectives (SLOs), %replace SLO
To improve throughput and system efficiency, existing LLM serving systems, such as Orca~\cite{yu_osdi2022_orca}, vLLM~\cite{kwon2023_vllm, kwon2023_pageattention}, and SGLang~\cite{zheng2023_sglang, zheng2024_sglang}, employ various optimization techniques, including continuous batching~\cite{yu_osdi2022_orca, kwon2023_vllm} and memory-efficient attention mechanisms such as PagedAttention~\cite{kwon2023_pageattention}. 
In particular, inspired by classical virtual memory and paging techniques, vLLM facilitates flexible sharing of KV (key-value) cache within and across requests to effectively reduce memory usage and thereby achieve high utilization in KV cache memory~\cite{kwon2023_vllm, kwon2023_pageattention}. 
These optimizations highlight the importance of memory management, especially in managing the KV cache, in achieving high system performance.
%, plays a vital role in boosting system performance and improving SLOs in LLM serving systems. 
}
%%
%%

%% ---- Figure 1: Memory layout (KV cache)  ----
\begin{figure}[t]
  \centering
  \includegraphics[width=1.0\columnwidth,trim={0 1.3cm 0 1cm},clip]{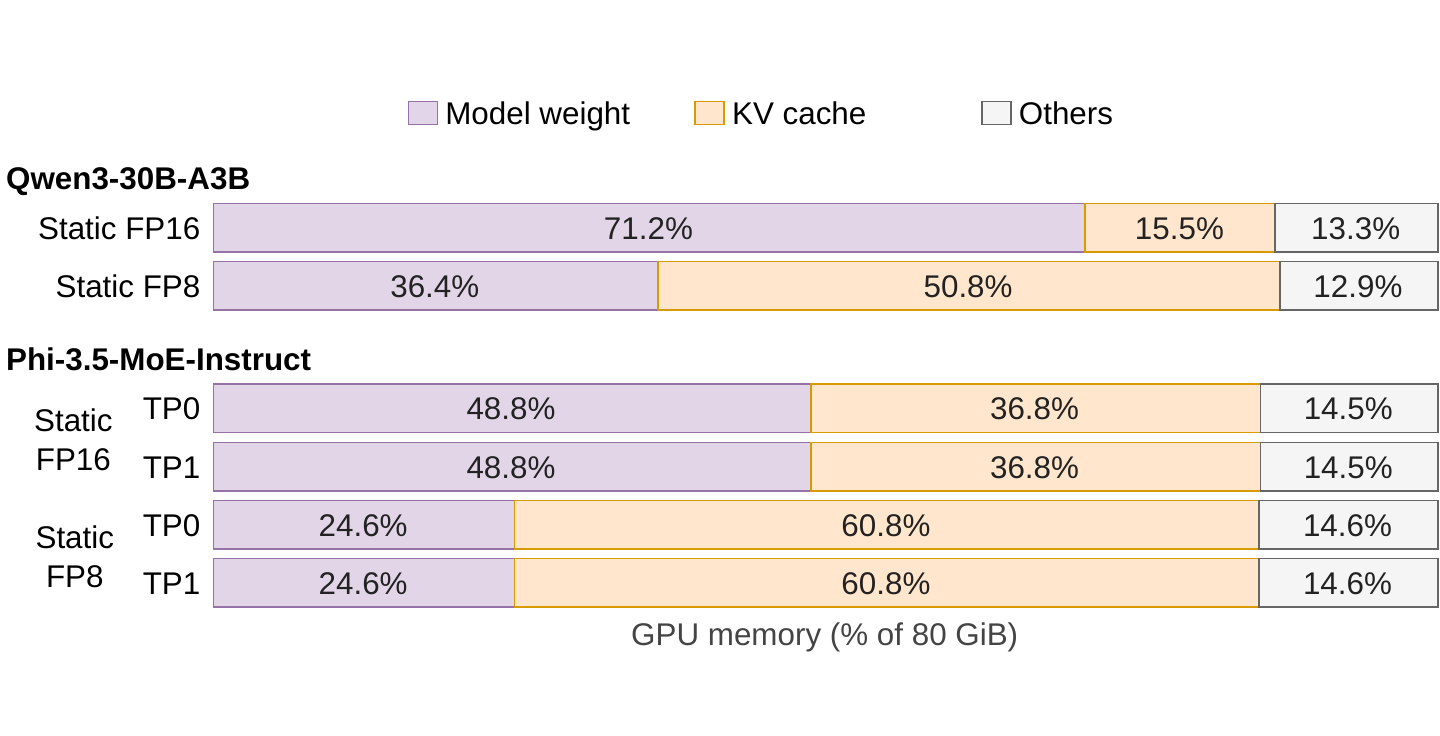}
  \caption{Memory layout when serving Qwen3-30B-A3B~\cite{yang2025qwen3technicalreport} and Phi-3.5-MoE-Instruct~\cite{abdin2024phi} on NVIDIA H100-80G GPUs~\cite{h100}.}
  %\caption{Memory layout when serving different LLMs on an NVIDIA H100-80G GPU. Accounting for around 30\%, the memory for the KV cache is (de)allocated per serving request. However, under high request rates, it may lead to frequent preemptions, thereby degrading overall throughput. \todo{Draw a new figure}}
  \label{fig:intro-layout}
\end{figure}

%% 
%Based on an auto-regressive Transformer model~\cite{vanswani_transformers}, modern LLMs mainly comprise \emph{self-attention (SA)} layers and multiple \emph{fully connected (FC)} layers, including query, key, and value generation, as well as projections. 
\dtt{
Modern LLM inference operates in two stages: a \emph{prefill} stage that processes the input prompt and a subsequent \emph{decode} stage that generates tokens.
The prefill stage is dominated by compute-intensive general matrix multiplication (GEMM) operations, which can be effectively accelerated on GPUs~\cite{v100, a100, h100} and Tensor Processing Units (TPUs)~\cite{tpu, tpuv4}.
In contrast, the decode stage is dominated by memory-intensive general matrix-vector multiplication (GEMV) or attention operations, which underutilize GPU compute resources.
LLM-serving systems such as Orca~\cite{yu_osdi2022_orca}, vLLM~\cite{kwon2023_vllm, kwon2023_pageattention}, and SGLang~\cite{zheng2023_sglang} enhance throughput by batching multiple requests together~\cite{yu_osdi2022_orca, kwon2023_vllm}, but the effective batch size is limited by the available GPU memory to store all the KV vectors of the active requests.}
%and PageAttention~\cite{kwon2023_pageattention}. 
%One common approach is to leverage more cost-effective solutions, such as gaming GPUs~\cite{rtx3060, rtx5060}, light-weight neural processing units~\cite{dfx, dfx_lpu}, or Processing-in-memory (PIM)~\cite{aimx, ianus_asplos2024, dear_pim}, to construct heterogeneous or disaggregated serving systems~\cite{ianus_asplos2024, splitwise}.
%}

\dtt{One fundamental challenge in LLM-serving systems is effectively managing GPU memory when dynamic workloads require different degrees of memory capacity for the KV caches~\cite{kwon2023_vllm}.
Popular LLM-serving systems such as vLLM~\cite{kwon2023_vllm}, SGLang~\cite{zheng2024_sglang}, TensorRT-LLM, TGI~\cite{huggingface_tgi}, llama.cpp, and DeepSpeed-Inference statically allocate memory for model weights and KV cache. Once the sizes of these memory spaces are determined, they will not be changed during runtime.} 
For example, Figure~\ref{fig:intro-layout} illustrates the memory distribution for a 30B-parameter LLM on an NVIDIA H100 GPU with 80GB RAM~\cite{h100}. 
71.2\% of the GPU memory is statically allocated to the model weights during serving. 
15.5\% of the memory is used to store the dynamic states of requests, which include the key and value tensors associated with the
attention mechanism, commonly referred to as KV cache.
Since model weights are assumed to be constant and activations occupy only a small fraction of GPU memory, how the KV cache is managed is critical to determining the maximum batch size~\cite{kwon2023_vllm}. 
When request traffic surpasses the system’s capacity, LLM serving systems like vLLM~\cite{kwon2023_vllm,kwon2023_pageattention} must prioritize a subset of requests and may need to preempt some requests, for example, the latest requests due to their first-come, first-served (FCFS) scheduling policy.

% Mention MoE models here. Larger model size, smaller KV cache size.

\dtt{
As a promising alternative to FP16, 8-bit floating-point formats (FP8) are gaining traction in LLM serving, offering up to 2× higher peak throughput and smaller memory footprint at a slight accuracy degradation~\cite{neurips22_fp8, kim2025_fp8_inference, shen2024_efficient_ptq_fp8, lee2026_nestedfp}. %For memory-intensive workloads, FP8 generally facilitates data transfer at 2× the rate of FP16, significantly boosting performance. 
%Storing and computing weights at FP8 are therefore potential methods to improve the computation utilization and reduce the memory required for model serving (???).}
%Additionally, FP8 support, particularly E5M2 (five exponent bits and two mantissa bits) and E4M3 (four exponent bits and three mantissa bits), is rapidly being adopted by modern hardware accelerators, including NVIDIA Hopper GPUs~\cite{h100} and Intel Gaudi HPUs.
 Interestingly, an FP8 model can be derived directly from the corresponding FP16 model and later restored to FP16 to recover accuracy~\cite{lee2026_nestedfp}. %, while parameters of most LLM models' layers, such as GPT~\cite{gpt3}, Llama~\cite{llama}, and Qwen~\cite{qwen1.5}, belong to a small range, for example, $[-1.75, 1.75]$, and can be transformed to FP8 (E4M3) with a small accuracy drop~\cite{lee2026_nestedfp}.
%More importantly, it is possible to revert from those FP8 models to FP16 models to fully restore the default accuracy mode. 
While conventional quantization methods do not allow efficient use of both high (unquantized) and low (quantized) precision, this property opens the opportunity of having an elastic KV cache memory: the memory occupied by the residual precision (the difference between FP8 and FP16) can be used as a shared resource between KV FP16 weights and the KV cache.}
\nxt{For instance, reconsidering the above case of serving a 30B model on an 80G GPU. With FP8, the model requires only 36.4\%, leaving 50.8\% for the KV cache, as illustrated in Figure~\ref{fig:intro-layout}. 
Similar patterns are also observed on Phi-3.5-MoE-Instruct ~~\cite{abdin2024phi} when it is served with two GPUs using tensor parallelism (TP). 
}

\dtt{We present~\sysname, a dual-precision serving mechanism for LLM serving systems to exploit this opportunity. 
Under normal loads,~\sysname serves the FP16-precision model (referred to as~\FullMode) with performance comparable to that of a conventional serving system.
When~\sysname identifies high KV pressure (request bursts or long-context inputs), 
~\sysname switches to the FP8 model (referred to as~\PerfMode), which is directly derived from the FP16 model, and dynamically reallocates the freed weight memory as additional KV blocks.
When KV pressure subsides,~\sysname restores the weights to gradually return the system to~\FullMode.} 

%This dynamically lowers the KV pressure, reducing the number of preempted requests and thereby improving system throughput at a small accuracy drop. 
%Meanwhile, under low KV pressure, the system restores the execution of the FP16 model to maintain the full accuracy mode. 
%Unfortunately, it is non-trivial to build such a mechanism and integrate it into existing serving systems, such as~\cite{kwon2023_vllm}. 
%}

%Inspired by the aforementioned observations, this work proposes~\sysname{}, a dual-mode precision LLM serving system with semi-unified memory, with the following main contributions.
The paper's contributions are summarized as follows.
\begin{itemize}
    \item \textbf{Semi-unified memory (SUM).} 
    %We introduce a novel semi-unified memory mechanism for an LLM serving system. 
    \dtt{
    We introduce a memory abstraction that partitions the original model weight memory into a persistent lower-precision region and a shared region. The shared region can store either the residual weights in~\FullMode or KV cache blocks in~\PerfMode. Our implementation is based on CUDA Virtual Memory Management to preserve compatibility with the default KV cache allocation in vLLM.
    }
    \item \textbf{Dual-precision execution with asymmetric switching.} 
    \dtt{
    DPS exploits the trade-off between FP8 and FP16 to balance accuracy and performance. Switching between these two precisions is asymmetric: transitioning from FP16 to FP8 requires no data movement, while the reverse requires Host-to-Device communication. We design a background restore mechanism that fully overlaps this communication cost with the ongoing FP8 computation, enabling smooth transitions.
    }
    \item \textbf{SUM-aware scheduling policy.} 
    \dtt{We integrate SUM with vLLM's scheduler. The scheduler actively monitors the KV cache pressure and determines when to switch to prevent cache thrashing and mitigate accuracy loss.
    }
    \item \textbf{Implementation and evaluation.} 
    \dtt{We implement DPS on top of vLLM and evaluate it across both dense and MoE models and various production workload traces. Our results show that \sysname improves sustained throughput by $2.1$--$3.3\times$ and effective pass@1 by up to $+41$\,pp over Static FP16, while preserving FP16-class accuracy.
    }
\end{itemize}

\section{Background}\label{sec:background}

\subsection{LLM architecture and LLM serving systems}~\label{subsec:LLMs and GPU-based LLMs Serving}
%\todo{Just take from UDP. Need to rephrase later.}
\dtt{
\textbf{LLM architecture.} Modern LLMs typically consist of stacked decoder blocks~\cite{gpt3, llama, touvron2023llama}. %, as illustrated in Figure~\ref{fig:llama3}. 
Each block is further decomposed into self-attention (SA) and feed-forward (FFN) layers.
Recent architectures introduce Mixture-of-Expert (MoE) layers~\cite{rajbhandari2022deepspeed}.
%normalization (LayerNorm~\cite{ba2016layernormalization} or RMSNorm~\cite{zhang2019root}), and residual connections~\cite{he2016deep}. 
The \textit{Self-Attention} module projects the previous layer's hidden states into query (Q), Key (K), and value (V) vectors and computes attention scores that represent each token's context awareness. 
%This mechanism helps the model dynamically focus on the relevant parts of the input sequence.
In autoregressive inference, to avoid redundant recomputation, self-attention relies on the K and V vectors computed from the previous steps. 
LLM inference systems generally store these KV vectors in GPU memory to reduce the load latency. 
%In attention-based autoregressive inference, it typically caches KV vectors in GPU memory (instead of CPU memory) to avoid recomputation and host-to-device load latency.
This memory cost becomes the performance bottleneck as the number of required KV vectors increases with the number of in-flight requests.
Insufficient KV cache capacity may require the system to use a high-latency memory storage or preempt requests. These activities degrade performance and may cause the system to run at low utilization or fail to meet service-level objectives (SLOs).
}

\begin{figure*}[t]
  \centering
  \includegraphics[
    width=1.4\columnwidth,
    trim={0 2.1cm 0 2cm},
    clip
  ]{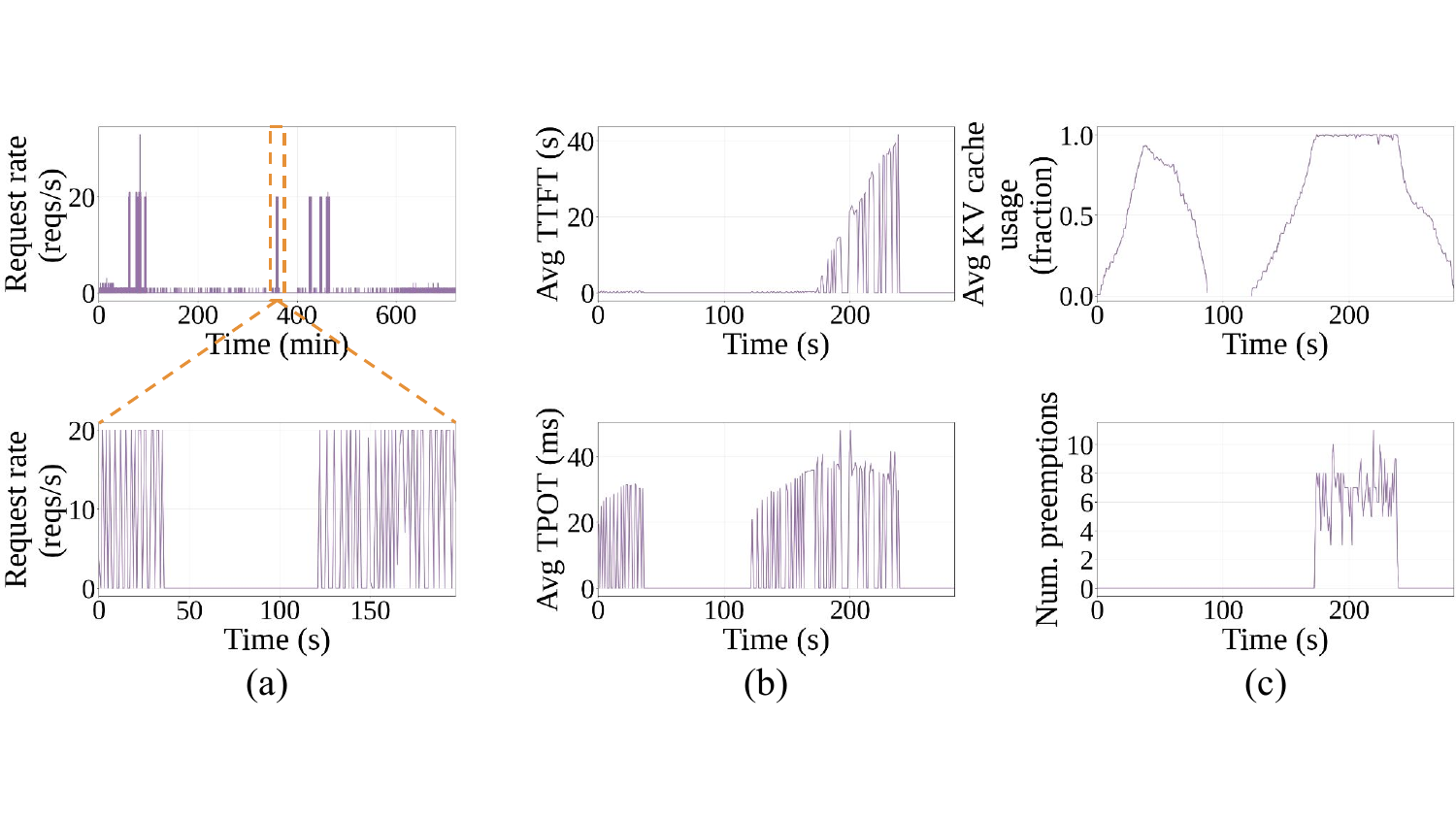}
  \caption{(a) Workload Fluctuations, (b) TPOT/TTFT, SLO (c) KV cache usage, preemptions. }
  \label{fig:motivation}
\end{figure*}

\noindent
\textbf{LLM serving systems.} 
LLM serving systems, such as vLLM~\cite{kwon2023_vllm} and SGLang~\cite{zheng2023_sglang}, focus on improving throughput and resource efficiency via various optimization techniques in scheduling and memory management. 
One essential optimization is continuous batching, which groups incoming requests into single batches to improve hardware utilization. 
vLLM introduces an advanced KV cache management algorithm, called PagedAttention or vLLM~\cite{kwon2023_vllm}. PagedAttention is inspired by virtual memory systems to enable flexible memory allocation and to share KV blocks between tokens and requests. These techniques also reduce memory fragmentation. SGLang uses prefix-aware caching and a reuse mechanism~\cite{zheng2023_sglang} to manage the KV cache.
Despite these advances, most systems statically partition the memory space between the weight and KV caches at initialization time.
Once allocated, the partitions remain fixed during the entire serving time. This design limits the flexibility to adjust the memory partitions in the presence of workload variation.

%A typical LLM serving stack is GPU-centric~\cite{a100, h100}, as illustrated in Fig.~\ref{fig:gpu}. Prior to LLM inference, the model weights are transferred from CPU memory to high-bandwidth memory (HBM) on a GPU device. During execution, tensors flow through the GPU memory hierarchy, from off-chip HBM and L2 cache to per-SM L1/shared memory, before being consumed by the compute units. Compute is split between CUDA cores, which provide general-purpose FP32/INT arithmetic, and Tensor Cores, which accelerate matrix math at much higher throughput. Across generations, Tensor Cores have scaled FLOP/s far faster than off-chip and on-chip bandwidths; e.g., on an A100-40GB the peak Tensor Core throughput dwarfs HBM bandwidth and both exceed host–device links by orders of magnitude~\cite{a100, dao2022flashattention}. Thus, in practice, LLM serving is often limited by data movement rather than math. Recent work drives Tensor Core utilization toward peak by: (1) \emph{reducing data movement and hiding memory latency} - tiling to on-chip SRAM and avoiding full attention materialization (FlashAttention~\cite{dao2022flashattention}), KV-cache paging and locality-aware schedulers (PagedAttention/vLLM~\cite{kwon2023_vllm}), quantization/compression to lower bytes-per-FLOP (GPTQ, AWQ, SmoothQuant~\cite{frantar2022gptq,lin2024awq,xiao2023smoothquant}); and (2) \emph{reducing non-Tensor Core overhead} - kernel/operator fusion~\cite{hsu2025ligerkernel}, persistent kernels/CUDA Graphs, and efficient hardware-aware dequantization~\cite{frantar2024marlin, lin2024qserve, zhang2024qqq}.

\subsection{FP16/FP8 for LLMs} 
%Unlike in computer vision tasks, where integer quantization is prevalent, LLMs often exhibit activation outliers, as demonstrated in various works~\cite{xiao2023smoothquant, lin2024awq, lin2024qserve, zhang2024qqq},. 
Unlike in computer vision tasks, LLMs often exhibit activation outliers~\cite{xiao2023smoothquant, lin2024awq, lin2024qserve, zhang2024qqq}, making floating-point formats (e.g., FP16 and FP8) widely used.  %Floating point formats such as FP16 or FP8 are commonly used in LLM serving. 
A floating-point format with $x$ exponent bits and $y$ mantissa bits is denoted as $ExMy$.
For instance, FP16 is also referred to as E5M10, where 5 and 10 bits are used for the exponent and mantissa, respectively. 
%For instance, the common half-precision float FP16 is also referred to as E5M10, where 5 and 10 bits are used for the exponent and mantissa, respectively. 
Two emerging FP8 formats, E5M2 and E4M3, are defined similarly. 
While FP16 has become the de facto standard for LLM serving,
FP8 is attracting significant attention with two fundamental factors: (1) the representational advantages of floating-point formats to capture outliers, and (2) the increasing hardware support for FP8 arithmetic.
Recent studies~\cite{neurips22_fp8, kim2025_fp8_inference, shen2024_efficient_ptq_fp8, lee2026_nestedfp} have shown that FP8 can maintain high accuracy with only small degradation compared to FP16. 
For example, FP8 (E5M2) has the same number of exponent bits as FP16 but lower precision because it has fewer mantissa bits. 
%As the data range is small, for example, the weight range of $[-1.75,1.75]$ in many layers of LLM models as reported in~\cite{lee2026_nestedfp}, FP8 (E4M3) still captures high dynamic ranges with four exponent bits, while offering a higher precision than FP8 (E5M2).
More importantly, FP8 reduces the bandwidth requirement by up to $2\times$, and FP8 is natively supported in arithmetic units by many modern hardware systems, including NVIDIA Hopper~\cite{h100}, Intel Gaudi HPUs, or NPUs~\cite{rngd_isscc25}.

\subsection{GPU Memory Characteristics}
\dtt{
Modern GPUs feature a hierarchical memory system, including high-bandwidth memory (HBM), on-chip caches (L1, L2, shared memory), and register files. 
The latency of accessing HBM is much higher than that of on-chip caches. However, on-chip cache sizes are much smaller.
Therefore, optimizing LLM operations mostly involves fitting the working set into on-chip and register memory.
%While GPUs offer high peak computational throughput, their performance is often constrained by memory bandwidth, especially for memory-bound workloads.
}
\dtt{
During LLM inference, the prefill phase benefits from compute-intensive GEMM operations that efficiently utilize NVIDIA Tensor Cores (Matrix Cores for AMD GPUs). 
However, the decode phase is dominated by memory-bound operations that require frequent HBM access. This results in low arithmetic intensity and GPU utilization.
}
\dtt{
As a result, optimizing memory usage and data movement is critical for improving overall system performance. 
In particular, reducing memory footprint or increasing effective memory capacity can directly translate into higher throughput.
}

% \begin{figure}[t]
%   \centering
%   \includegraphics[
%     width=1.0\columnwidth,
%     trim={0 2.4cm 0 4.5cm},
%     clip
%   ]{figure/fig2.pdf}
%   \caption{(a) Workload Fluctuations, (b) TPOT/TTFT, SLO (c) KV cache usage, preemptions. }
%   \label{fig:motivation}
% \end{figure}
\section{Motivation}\label{sec:motivation}

% \begin{figure*}[t]
%   \centering
%   \includegraphics[width=1.5\columnwidth]{dummy/motivation_from_morphServe.PNG}
%   \caption{(a) Workload Fluctuations, (b) TPOT/TTFT, SLO (c) KV cache usage, preemptions.}
%   \label{fig:motivation}
% \end{figure*}

\subsection{Challenge: Dynamic Load in LLM Serving}
\noindent
\textbf{Bursty LLM workloads.} LLM serving systems face highly dynamic and bursty computing patterns due to substantial variability in the request arrival rates (i.e., request bursts), input prompt, and output sequence lengths across requests~\cite{sharegpt, kdd2025_burstgpt}.
As illustrated in Figure~\ref{fig:motivation}(a), the production workloads of the Microsoft Azure LLM services and BurstGPT reveal rapid fluctuations: the request rate exhibits nearly a fivefold variation between the lowest and highest load periods.
In addition, these fluctuations occur with high temporal frequency, indicating second-level variability.
These fluctuations pose a practical challenge for real-world LLM inference workloads, which are not fully captured in settings such as vLLM~\cite{kwon2023_vllm}.
\dtt{
An important observation is that the periods of high KV cache pressure are not permanent; they are usually interleaved with low-pressure periods.
This transience has two implications. First, static solutions that constantly trade quality for performance (e.g., always using FP8 or static weight offloading), or vice versa, are suboptimal because they incur a permanent cost for an intermittent problem.
Second, rapid fluctuations demand a lightweight switching mechanism, which in turn requires an efficient system design.
This observation motivates a fast, runtime-adaptive memory management mechanism that can elastically respond to different workload characteristics.
}

\noindent
\textbf{Long TTFT/TPOT and SLO violation.}
A request burst may cause a long Time-to-First-Token (TTFT) latency and SLO violations, as illustrated in Figure~\ref{fig:motivation}(b).
As the system load increases, even small surges can cause sharp spikes in TTFT latency.
Specifically, a serving system quickly exceeds the SLO threshold once GPU memory becomes insufficient to schedule new requests for prefilling or to continue decoding for the ongoing batch. 
Consequently, an incoming request is forced to wait until memory is reclaimed, incurring
significant queuing latency with SLO violation.
\dtt{
The problem is exacerbated by long-context requests, especially in modern applications, where the context length can reach 32K-1M tokens~\cite{rando2025longcodebenchevaluatingcodingllms, jimenez2024swebenchlanguagemodelsresolve, cao2026qwen3codernexttechnicalreport}.
}

% \begin{figure*}[t]
%   \centering
%   \includegraphics[
%     width=1.5\columnwidth,
%     trim={0 2.1cm 0 2cm},
%     clip
%   ]{figure/fig2.pdf}
%   \caption{(a) Workload Fluctuations, (b) TPOT/TTFT, SLO (c) KV cache usage, preemptions. }
%   \label{fig:motivation}
% \end{figure*}

\noindent
\textbf{KV Cache Pressure and Preemptions.}
The above issue can also be visualized via KV cache utilization or pressure. 
As the system load increases, the KV cache memory is highly utilized. 
When the request traffic surpasses the system’s capacity, the serving systems like vLLM~\cite{kwon2023_vllm, kwon2023_pageattention} must prioritize a subset of requests and need to preempt some requests. 
As illustrated in Figure~\ref{fig:motivation}(c), when the system load increases with high fluctuations, the KV cache memory is nearly full, and preemptions occur more frequently. 

\dtt{
Preemptions are particularly costly for long-context workloads. A preempted request incurs recomputation of the entire KV cache from scratch at the rescheduling point, or swapping all relevant KV cache items between GPU and CPU memory.
Either way, the cost is proportional to the input context length. For a coding agent processing a 64K-token repository context, preemption can add seconds of latency that generally exceed SLO budgets.
Lastly, preemption prevents current LLM-serving systems such as vLLM~\cite{kwon2023_vllm} and SGLang~\cite{zheng2024_sglang} from fully utilizing GPUs.
As the context length continues to grow, driven by both user demand and model support, the preemption cost increases proportionally, making preemption-avoidance solutions more attractive for maintaining the SLO.
%As a result, it may incur queuing latency with an SLO violation because an incoming request is forced to wait until memory is reclaimed.
}

% \begin{figure}[t]
%   \centering
%   \includegraphics[width=\columnwidth]{figure/fig_sum.png}
%   %\caption{\todo{Draw a new figure: (a) Conventional layout, (b) New layout with SUM for dual-mode FP16/FP8 serving.}}
%   \caption{SUM}
%   \label{fig:intro-malloc}
% \end{figure}

% \begin{figure*}[t]
%   \centering
%   \includegraphics[width=2\columnwidth]{figure/DPS.drawio.pdf}
%   \caption{Dual-mode precision execution and shared-region dynamics in DPS. (a) Dual-precision execution: FP16 weights are decomposed into an FP8 tensor and a residual tensor that together reconstruct the original value during full-precision execution, while performance mode execution uses only the FP8 tensor. (b) Weight-to-KV transition: when the free pool is exhausted, the SUM Manager satisfies KV block allocation requests by drawing from residual-backed units, marking the affected residual mappings as stale, and switching the system to performance mode in which residual tensor data are not read. (c) KV-to-weight transition: when KV blocks are freed, the SUM Manager wakes a background restore worker that rebinds residual tensors to newly freed units and populates them from the CPU backup, returning the system to full-precision mode once all residual tensors are restored.}
%   \label{fig:shared_tegion}
% \end{figure*}

\subsection{\nxt{KV Cache Pressure and Preemption Mitigation via Dual-Precision Serving (DPS)}}
\noindent
\nxt{\textbf{Opportunity.}
\emph{Dual-precision serving} presents an opportunity to mitigate KV cache pressure. 
For example, consider a memory layout for serving a 13B model on an A100-40G GPU. 
When applying a nested FP8/FP16 model~\cite{lee2026_nestedfp}, 13 GB, 13 GB, and 12 GB are primarily allocated for upper and lower FP8 weights and KV cache, respectively, accounting for 95\% of the GPU memory. 
However, under a large number of requests, the serving system performs a~\PerfMode{} that uses only 13 GB of upper FP8 weights for computation, thereby improving system performance. 
We observe that 13 GB of the lower FP8 weights can be deallocated and remapped to the KV cache.
As a result, in this case, the KV cache can be elastically extended by 108.33\% (= 13/12), significantly mitigating KV cache pressure and improving system throughput and SLOs.
Meanwhile, when all 26 GB are used for the upper and lower FP8 weights, this preserves the behavior of an LLM serving system, such as vLLM~\cite{kwon2023_vllm}, under a~\FullMode{}.}

\noindent
\nxt{\textbf{Challenge.}
The above \emph{dual-precision serving} presents several challenges. 
Firstly, serving an LLM in~\FullMode{} with both upper and lower weights should incur negligible overhead compared with a baseline system, such as~\cite{kwon2023_vllm}. 
Secondly, extending the KV cache size, for example, from 12GB to 25GB, must preserve the powerful page KV cache management in existing LLM serving systems. 
This requires a~\emph{lightweight accuracy-to-performance} mode switching and a virtual memory scheme to elastically extend the KV cache. 
Lastly, restoring a~\FullMode{} from a~\PerfMode{} intuitively requires loading 13 GB of lower FP8 weights from the host to GPU memory, which may cause considerable latency during serving. 
This poses a critical challenge for designing an effective mechanism for~\emph{smooth fast-to-full} mode switching.
}

\section{Method}\label{sec:method}

% \begin{figure}[t!]
%   \centering
%   \subfloat[Structure of a Llama-3 model]{%
%     \includegraphics[width=\columnwidth]{figure/fig4_sum_layout_v7_white.pdf}
%     \label{fig:llama3}
%   }\\[0.5em]
%   \subfloat[GPU memory hierarchy and data flow to Tensor/CUDA cores]{%
%     \includegraphics[width=\columnwidth]{figure/fig5_sum_system_architecture__1__white.pdf}
%     \label{fig:gpu}
%   }
% \caption{(a) Conventional layout, (b) New layout with SUM for dual-mode FP16/FP8 serving.}
% \label{fig:sum-abstraction}
% \end{figure}

%% NXT ++ 2026.04.27
%% Flow
%% 1. Overview, Scope, and Assumptions: 
%%  - Nested or Unified models: M_0 < M_1 < ... < M_N
%%      + Example: Special case: N = 1 --> Two models like Nested FP, where M0 an M1 represent FP8 and FP16 models
%%      + Without loss of generality, focus on N=1: 
%%  - Key properties
%%      + Bounds: 
%%          - For model parameters: 
%%          - For KV cache: Limit vs. preemptions
%%      + Run-time Switching
%%      + KV Reallocation: Page-based management
%%
%% 2. Nested-Model Serving
%%
%% 3. System Integration

%\subsection{Overview}
\subsection{\nxt{Design Methodology}}
% DPS is built around a single design principle: weight memory should be elastic. 

\begin{figure}[t]
  \centering
  \includegraphics[width=1\columnwidth]{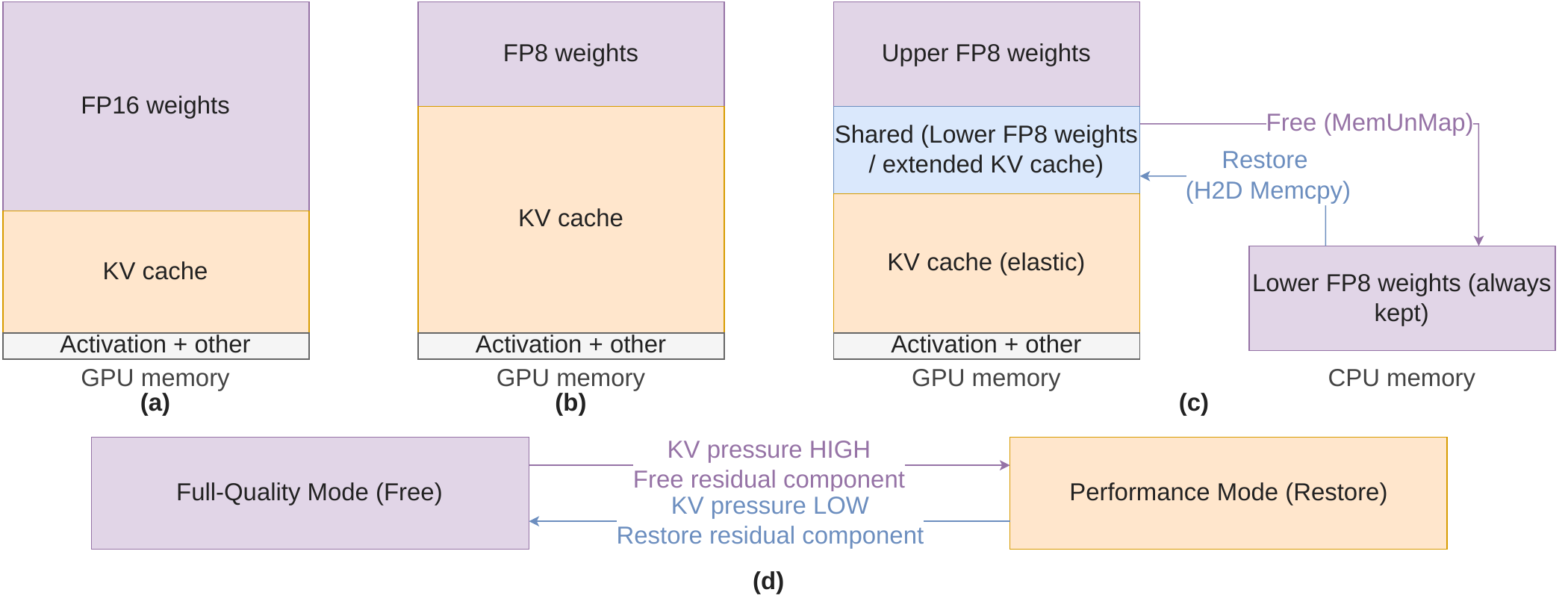}
  %\caption{\todo{Draw a new figure: (a) Conventional layout, (b) New layout with SUM for dual-mode FP16/FP8 serving.}}
  %\caption{Static memory layout (left) vs Semi-unified memory (right). \todo{Redraw and simplify the figure with key messages (Truong): (1) Persistent weight vs. page-based KV cache (2) Base model for performance: Persistent, and a semi-unified memory (SUM) shared by page-based KV cache and a layer-based residual model.}}
  \caption{\nxt{Three memory layouts: (a) FP16 model, (b) FP8 model, and (c) proposed semi-unified memory (SUM) with dynamic switching between a~\FullMode{} and a~\PerfMode{} (d).}}
  %SUM consists of a persistent region for an FP8 model and a shared region that is dynamically mapped to either a residual model or KV blocks. With SUM, under high KV pressure, the system quickly switches from a~\FullMode{} to a~\PerfMode{} by executing a persistent FP8 model, thereby enabling a larger KV cache to boost performance. Meanwhile, under low KV pressure, the system smoothly recovers to the~\FullMode{} by reloading residual memory via a host-to-device (H2D) memory copy.}}
  \label{fig:mem-layout}
\end{figure}

\noindent
\nxt{\textbf{Persistent and Dynamic Memory.} 
This subsection revisits the fundamental concept of memory management in LLM-serving systems, such as vLLM~\cite{kwon2023_pageattention}. 
Specifically, inspired by a traditional virtual memory mechanism in an operating system, KV blocks are typically managed in pages and dynamically (de)allocated during serving. 
Notably, a model's parameters are typically assumed to be preloaded to GPU memory before serving and considered persistent during serving. 
For example, FP16 and FP8 models are loaded into a persistent region, as illustrated in Figures~\ref{fig:mem-layout}(a) and (b), respectively. 
This assumption is intuitive because a typical LLM model is large, making it costly to transfer an entire model from CPU memory to GPU memory at runtime. 
Meanwhile, a fundamental observation is that if LLM inference can be performed with a partial model during serving, KV memory can be temporarily extended, possibly relaxing KV memory pressure and thereby improving system throughput. 
Intuitively, instead of being persistent, weight memory may be elastic, with parts of it mapped to KV blocks at runtime. 
}

\noindent
\nxt{\textbf{Dynamic Networks and a Nested Model.} Under bursty workloads, KV cache pressure and a demand for a KV cache extension may occur instantly, which becomes difficult to predict. 
To effectively handle such an instant KV pressure, a system must quickly switch from a persistent weight memory to a temporarily elastic one, leaving more space for KV blocks. 
Dynamic networks, including nested FP~\cite{lee2026_nestedfp}, any precision networks~\cite{yu_aaai_2021_dynamic, icml24_anyprecision}, early-exit networks~\cite{Jeon_2024_WACV_ee, chen_icml_2024_ee, xu_isca_2025_ee}, provide a promising opportunity to address the challenge. 
Without loss of generality, let's consider an FP16 model (e.g., E5M10) storing both an FP8 model (e.g., E5M2 or a custom E4M3 format~\cite{lee2026_nestedfp}) and a residual model, as illustrated in Figure~\ref{fig:mem-layout}(c).
More specifically, an FP16 model is split into two parts: an FP8 model (e.g., with upper (or MSB) weight tensors) stored in a persistent region and a residual model (e.g., with lower (or LSB) weight tensors) stored in a shared region. 
This can facilitate quick switching between FP16 and FP8 inference, temporarily freeing the residual model's memory for KV blocks and thereby boosting system performance. 
}

\noindent

\begin{figure*}[t]
  \centering
  \includegraphics[width=2.05\columnwidth]{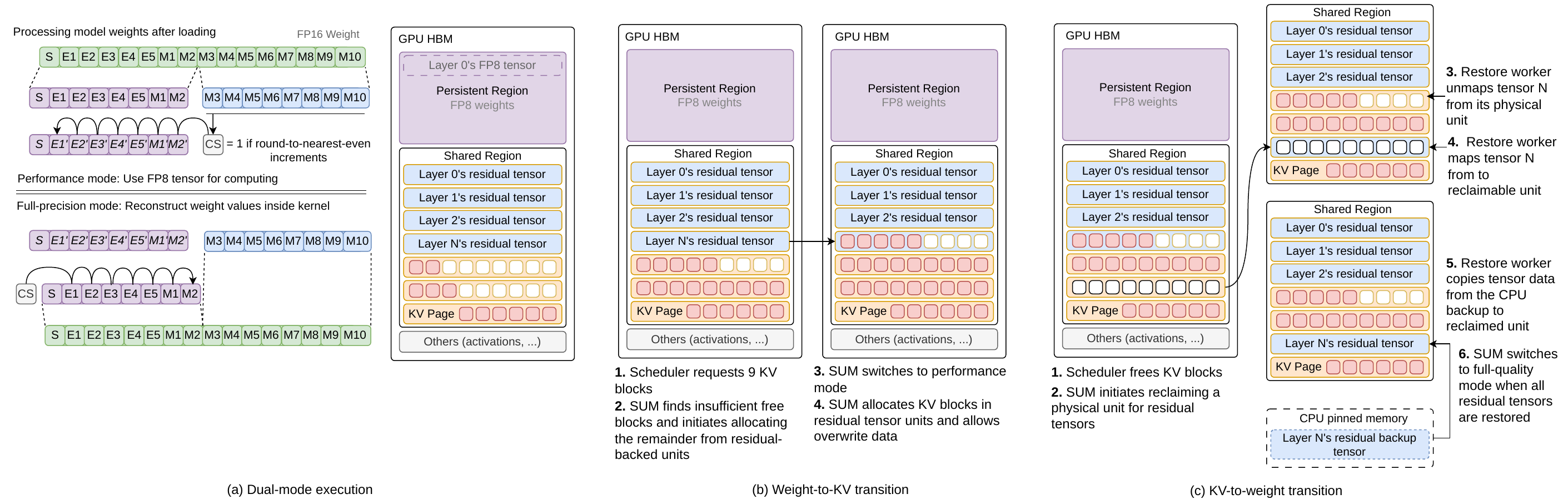}
  \caption{DPS with SUM: (a) SUM mapping for nested FP16-FP8 serving, (b) Weight-to-KV transition, and (c) KV-to-weight transition.}
  %\caption{Dual-mode precision execution and shared-region dynamics in DPS. (a) Dual-precision execution: FP16 weights are decomposed into an FP8 tensor and a residual tensor that together reconstruct the original value during full-precision execution, while performance mode execution uses only the FP8 tensor. (b) Weight-to-KV transition: when the free pool is exhausted, the SUM Manager satisfies KV block allocation requests by drawing from residual-backed units, marking the affected residual mappings as stale, and switching the system to performance mode in which residual tensor data are not read. (c) KV-to-weight transition: when KV blocks are freed, the SUM Manager wakes a background restore worker that rebinds residual tensors to newly freed units and populates them from the CPU backup, returning the system to full-precision mode once all residual tensors are restored.}
  \label{fig:shared_region}
\end{figure*}

\noindent
\nxt{\textbf{Dual-Precision Serving (DPS) Overview.} 
Inspired by the observations above, this subsection presents an overview of DPS, a dual-precision serving mechanism that can be directly integrated into existing LLM serving systems, such as vLLM~\cite{kwon2023_pageattention}. 
DPS introduces a novel semi-unified memory (SUM) scheme.  
SUM is a memory abstraction that allows a shared physical region to dynamically back two distinct virtual tensors: one for the residual weights, the other for the KV cache pool. 
In the following subsection, we first discuss the SUM architecture and asymmetric mode switching. 
Next, dual-precision execution is explained in detail. 
Lastly, we provide a theoretical performance bound for a DPS problem. 
}

\subsection{\nxt{Mapper: Semi-Unified Memory Abstraction}}
SUM is composed of four main components: a persistent region holding the lower-precision representation of model weights, a shared region whose physical memory may back either residual high-precision weights or KV cache blocks, a CPU-side residual backup used for restoring residual weight data after it has been overwritten, and a coordinator, called the SUM Manager, that owns the binding state of the shared region and exposes allocation operations to the serving system.

\noindent
\textbf{Persistent Region.} 
The persistent region holds the lower-precision representation of model weights. 
Specifically, the region can be used to store an FP8 model, as illustrated in Figure~\ref{fig:shared_region}(a). 
It is allocated via the standard PyTorch allocator at deployment startup and occupies a fixed portion of GPU memory throughout the deployment's lifetime. 
\remark{Notably, this retains the structure of the conventional serving systems like vLLM~\cite{kwon2023_pageattention}:~\emph{the system can always execute a model, for example, an FP8 one, regardless of the state of the shared region or KV cache}. 
For an FP8 model, one can directly obtain an FP8 tensor in E5M2 format from its corresponding FP16 (E5M10) tensor by using all five exponent bits and keeping the two MSB bits in the mantissas. 
In a specific case, if the range of data is small and the first MSB bit in the exponent is zero, one can obtain an FP8 tensor with an E4M3 format by keeping four out of five exponent bits and three MSB mantissa bits, as illustrated in Figure~\ref{fig:shared_region}(a). 
This has been exploited in~\cite{lee2026_nestedfp}, as many layers in LLMs have a small weight range. 
}
%\remark{Specifically, the region can be used to store an FP8 model, as illustrated in Figure~\ref{fig:shared_region}. 
%Notably, an FP8 tensor can be obtained in different ways. 
%For instance, an FP8 (E5M2) can be directly obtained from its corresponding F16 (E5M10) one. 
%Furthermore, if data is in a small range (e.g., when the MSB bit E1 in the exponents is zero), an FP8 (E4M3) can be also obtained by ignoring the MSB bit in the exponents~\cite{lee2026_nestedfp}, as shown in Figure~\ref{fig:shared_region}.
%}
%It is allocated through the standard PyTorch allocator at deployment startup and occupies a fixed portion of GPU memory throughout the lifetime of the deployment. 
%It guarantees that the system can execute the lower-precision model at any point during operation, regardless of the state of the shared region.

%\subsubsection{Shared Region}
\noindent
\textbf{Shared Region.} 
The shared region is the elastic component of SUM. 
It consists of two primitives: \textit{physical units} and \textit{virtual tensors}.
A physical unit is a contiguous region of GPU memory configured via the CUDA Virtual Memory Management (VMM) API~\cite{nvidia_vmm_api, nvidia_vmm}. 
A virtual tensor is a contiguous virtual address range exposed to the serving system through standard tensor access patterns. 
CUDA VMM separates virtual address reservation, physical handle allocation, and mapping between them, allowing a single physical handle to be mapped into multiple virtual addresses simultaneously and remapped at runtime. 
SUM relies on this separation to maintain dynamic bindings between physical units and virtual tensors.

\remark{It is essentially noted that SUM, in particular the shared region, builds a bridge between a~\emph{persistent} model and a dynamic~\emph{page-based KV cache} memory in a serving system, such as vLLM~\cite{kwon2023_pageattention}, as demonstrated in Figure~\ref{fig:shared_region}.  
When the physical shared memory is entirely allocated to page-based (virtual) KV blocks, this facilitates serving an FP8 model stored in the persistent region. 
This refers to a~\PerfMode{} in which the KV cache can fully utilize shared memory, thereby possibly reducing KV cache pressure and preemptions and boosting performance. 
It can also increase system throughput by leveraging FP8 computing, e.g., halving memory accesses by 2x or doubling peak tensor-core throughput with hardware support~\cite{lee2026_nestedfp, h100}. 
Meanwhile, the shared region can be overlaid and assigned to a partial persistent model. 
For example, as described in Figure~\ref{fig:shared_region}(a), the shared region can be configured to store a residual FP8 model that is combined with the FP8 model in the persistent region to form an FP16 model. 
This enables a~\FullMode{} because the entire FP16 model becomes persistent in GPU memory, as in conventional vLLM~\cite{kwon2023_pageattention} with FP16 model execution. 
Essentially, SUM provides a wrapper for a~\emph{persistent} model and a dynamic~\emph{page-based KV cache} in a serving system, while retaining effective KV cache management as in existing serving systems, such as vLLM~\cite{kwon2023_pageattention}.
}
%Each physical unit is sized to the residual weight tensor of a single transformer layer. ...(???)

%\subsubsection{Residual Backup} 

\noindent
\textbf{Residual Backup.} 
The residual backup is a pinned, CPU-side copy of all transformer layers' residual weight tensors, initialized once at deployment startup. 
It serves as the source for restoring residual weights after physical units have been overwritten by KV cache writes. 
\remark{More specifically, after the physical shared memory is entirely assigned to page-based (virtual) KV blocks, the residual weights in GPU memory are lost and overwritten by KV. 
When KV blocks are freed, leaving sufficient space in the shared region for a residual weight, the residual backup provides a source for reloading residual weights from CPU into the shared region on GPU.
}

\noindent
\textbf{SUM manager.} 
The SUM Manager maintains the binding state of all physical units in the shared region and records the current mapping between physical units and virtual tensors. 
It processes allocation and deallocation requests for KV cache blocks issued by the serving system. 
When an allocation request targets units that hold no residual data, the SUM Manager returns the requested blocks directly. 
When a request targets a unit holding a residual-weight virtual tensor, the SUM Manager signals the precision-mode transition to the model executor and marks the affected residual mapping as stale.

The SUM Manager observes deallocation events from the serving system to identify physical units whose KV cache blocks have all been freed. It enqueues a restore task for each such unit, and a background worker thread internal to the SUM Manager processes the queue: for each task, the worker selects a stale residual weight virtual tensor, unmaps it from its previous unit, maps it to the reclaimable unit, and copies the layer's residual weight data from the residual backup into the unit. 
Once all layers' residual weight data have been restored, the SUM Manager signals the model executor to transition back to full-precision execution. Performing restore work in the background keeps VMM operation latency off the allocation and deallocation hot path.

\noindent
\textbf{Dual-mapping property.} 
An alternative design would unmap each physical unit before rebinding it to a different virtual tensor, avoiding the dual-mapping property entirely. Such a design is incompatible with vLLM's asynchronous scheduling model. 
In this model, the scheduler may mark KV blocks as free while GPU workers still hold in-flight reads issued against those blocks by earlier submitted work. 
If freeing a KV block involved unmapping its backing unit, the in-flight reads would access unmapped memory, causing an illegal memory access error. 
Resolving this race would require synchronizing the scheduler against the workers on every free event, which defeats the purpose of asynchronous scheduling. SUM avoids the race by construction: the KV virtual tensor is permanently mapped to every physical unit in the shared region, so marking a unit's residual mapping as stale does not affect the KV mapping, and in-flight reads through the KV path remain valid regardless of residual state.

\subsection{Scheduler: Dual-Precision Execution}
\begin{figure}[t]
  \centering
  \includegraphics[width=1.05\columnwidth]{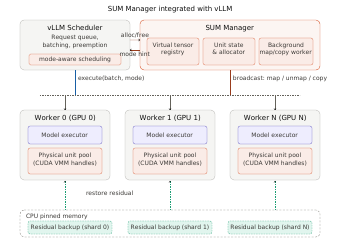}
  %\caption{\todo{Draw a new figure: (a) Conventional layout, (b) New layout with SUM for dual-mode FP16/FP8 serving.}}
  \caption{Dual-precision serving architecture with SUM backend.}
  \label{fig:system-arch}
\end{figure}

\noindent
\remark{\textbf{Overall concept.} 
SUM facilitates non-stop execution and can be integrated into vLLM, as illustrated in Figure~\ref{fig:system-arch}. 
Specifically, at runtime, the model executor follows one of two paths depending on the chosen mode. 
In a~\FullMode{}, it combines the FP8 weight tensor from the persistent region with the residual weight tensor from the shared region to reconstruct the original FP16 values during computation. 
In a~\PerfMode{}, only the FP8 weight tensor is used; the residual weight tensor in the shared region is not accessed, and its physical memory may at this point be holding KV cache data rather than residual weight data. 
The SUM Manager assigns a current precision mode to each worker, and the model executor reads that mode at the start of each forward pass. 
This mode is set to a~\PerfMode{} when the SUM Manager processes an allocation request that overwrites a residual-bound unit, and reset to a~\FullMode{} when the background restore worker has populated all residual weight tensors. 
Mode transitions are batch-aligned: the SUM Manager updates the mode only between forward passes, and the model executor executes each pass entirely in a single mode. 
Obviously, SUM also enables single-mode execution. 
}

\noindent
\remark{\textbf{Dual-mode serving (DPS) with asymmetric switching.} 
This subsection unleashes the power of SUM for dual-precision serving. 
Notably, runtime switching for LLM serving with multi-precision representations has recently been discussed~\cite{lee2026_nestedfp, icml24_anyprecision, matgtq}. 
DPS implements GroupedGEMM kernels that perform NestedFP-style reconstruction~\cite{lee2026_nestedfp,neurips22_fp8} and support dense LLMs and MoE models. 
DPS extends the encoding scheme with an alternative FP8 E5M2 representation for layers whose weight magnitudes fall outside the range that E4M3 can express. 
With this extension, all projection layers in the model are eligible for FP8 execution, and the per-layer encoding is selected offline at model loading time based on weight statistics. 
Inspired by~\cite{lee2026_nestedfp}, SUM facilitates a nearly-zero-cost switching from a~\FullMode{} to a~\PerfMode{}, as described in Figure~\ref{fig:shared_region}(b). 
Specifically, because the FP8 model is in the persistent region, the system can execute it directly without stalling while loading a new model.
It is especially effective for bursty workloads because the demand for a~\PerfMode{} appears instantly and non-deterministically. 
More importantly, unlike~\cite{lee2026_nestedfp}, the shared memory for the (previously persistent) residual weight is unmapped and reallocated to KV cache blocks, effectively mitigating high KV cache pressure and potentially reducing preemptions. 
}

%% Merge from a Tuan's
\remark{SUM also facilitates a restoration of a~\FullMode{}, as illustrated in Figure~\ref{fig:shared_region}(c). 
Restoration is the process of bringing the residual weight (the lower FP8 byte) back from CPU memory to GPU memory so that subsequent forward passes can run in a~\FullMode{}.
We use a dedicated non-blocking copy stream to perform the required Host-to-device (H2D) memory copy and to overlap it with the ongoing computation.
Since GPUs generally have a single PCIe copy engine, long H2D copies can stall the entire decoding pipeline. 
To avoid this problem, we schedule copies in small chunks and let other system H2D copies run between chunk copies. 
This fragmentation deliberately leaves the engine between chunks so that small H2D traffic from the request scheduler can interleave without significant delay.
At engine startup, when many copies are issued back-to-back and no other PCIe traffic is present, the system uses a simpler one-shot path: each copy is a single asynchronous transfer, terminated by a CUDA event that the compute stream waits on. 
Chunking is only reserved for the runtime restore path.
Empirically, we observe that a fixed chunk size of 4MiB works well for most cases.
Notably, because H2D transfers overlap with ongoing inference and there is no synchronous wait that pauses the engine, the actual cost paid is therefore not the duration of the transfer but the slight PCIe contention it imposes on per-step scheduler uploads, which the chunked yield design is meant to bound. 
}
%More importantly, this only delays a transition from a~\PerfMode{} to a~\FullMode{}, without interrupting the execution of the current~\PerfMode{} during this transition. 
%}

\noindent
\textbf{Granularity of the precision switch.}
Notably, restoration cannot itself preempt or replay a request; once a request's decode has produced a token under single-weight precision, the token is final, and only subsequent decode steps are affected by the restored precision.
The system pre-captures two CUDA graphs per supported batch size: one with single-weight (FP8-only) and the other with dual-weight (FP8+residual) kernels. The dispatcher picks exactly one for each forward pass. 
As a result, a precision change takes effect at scheduling-step boundaries, never within a step: every layer in a given forward pass runs at the same precision. 
A request that spans a mode transition continues without a restart: tokens generated before the flip use FP8, and subsequent tokens use FP16, but the request itself is never recomputed.

\begin{algorithm}[t]
\caption{KV block allocation and deallocation in SUM}
\label{alg:alloc}
\begin{algorithmic}[1]
%\small % Changes body text to small
\footnotesize % Changes body text to tiny
\State $\textsc{Free}$: units w/ no KV blocks and no residual mapping.
\State $\textsc{Backed}$: units bound to residual weight virtual tensors
\State $\textsc{Avail}$, $\textsc{Full}$: KV units with at least one free block or with no free blocks
%\State $\textsc{Full}$: KV units with no free blocks
\State $\textsc{Stale}$: invalid residual mappings awaiting restoration
\State $\textsc{Evicted}$: weights whose residual data is on CPU only, awaiting restoration
\Procedure{AllocKVBlock}{$n$}
    \State $\text{result} \gets \emptyset$
    \While{$|\text{result}| < n$}
        \If{AVAIL is empty}
            \If{FREE is non-empty}
                \State $u \gets$ pop from FREE
            \ElsIf{BACKED is non-empty}
                \State $u \gets$ pop from BACKED
                \State activate lower-precision mode
                \State mark $u$'s residual mapping as stale
            \Else
                \State \Return failure
            \EndIf
            \State add $u$ to AVAIL
        \EndIf
        \State $u \gets$ pop from AVAIL
        \State allocate up to $n - |\text{result}|$ KV blocks from $u$; append to result
        \If{$u$ has no free blocks} add $u$ to FULL \Else{} add $u$ to AVAIL \EndIf
    \EndWhile
    \State \Return result
\EndProcedure

\Statex

\Procedure{FreeKVBlock}{$\text{blocks}$}
    \For{each unit $u$ containing a block in blocks}
        \State mark the corresponding blocks free in $u$
        \If{$u$ was in FULL} move $u$ to AVAIL \EndIf
        \If{$u$ has no allocated blocks}
            \State remove $u$ from AVAIL
            \State add $u$ to FREE
            \If{EVICTED is non-empty}
                \State wake \textsc{RestoreWeights}
            \EndIf
        \EndIf
    \EndFor
\EndProcedure
\end{algorithmic}
\end{algorithm}

\subsection{KV-Cache-Pressure-aware DPS} \label{subsec:KV-aware}
\noindent
%\subsubsection{Precision-switching Algorithm}
\remark{\textbf{Precision-switching Algorithm.}
This subsection presents a DPS algorithm based on KV cache pressure. 
The central idea is to build a KV block allocation and deallocation/free with SUM, as summarized in Algorithm~\ref{alg:alloc}. 
This is based on runtime internal states of SUM, including $\textsc{Free}$, $\textsc{Backed}$, $\textsc{Avail}$, $\textsc{Full}$, and  $\textsc{Stale}$. 
Specifically, $\textsc{Free}$ stores \emph{free} physical units that are not assigned to either a KV block or a residual weight block. 
$\textsc{Backed}$ saves units bound to residual weight virtual tensors. 
While $\textsc{Avail}$ marks KV units with at least one free block, $\textsc{Full}$ indicates KV units with no free blocks.
$\textsc{Stale}$ marks invalid residual mappings awaiting restoration. 
}

\remark{
The KV block allocation is presented in lines 6-28 of Algorithm~\ref{alg:alloc}. 
Consider a request to allocate $n$ new KV blocks and let $u$ be a temporary list to store physical units for these KV blocks. 
The procedure starts by checking if $\textsc{Avail}$ is empty. 
If there are no KV units with at least one free block (line 9), it first tries to find an empty allocation slot in $\textsc{Free}$ (lines 10-11).
Notably, this step retains effective KV management in a serving system such as vLLM~\cite{kwon2023_pageattention}. 
The key idea is that if an unused slot exists in the KV cache, it reserves it for a new KV block by updating the allocation slot list $u$. 
Meanwhile, if there is no available slot and this is under a~\FullMode{} (e.g., non-empty $\textsc{Backed}$) (line 12), it updates $u$ from $\textsc{Backed}$ (line 13) and marks the residual mapping of $u$ as stale (line 15). 
Also, it activates a~\PerfMode{} (line 14). 
This step clearly demonstrates that~\sysname{} can instantly switch from a~\FullMode{} to a~\PerfMode{} to resolve KV cache pressure (e.g., empty $\textsc{Free}$). 
Lastly, the allocation slot list $u$ is used to allocate KV blocks (lines 21-25).
}

\remark{The KV block deallocation is shown in lines 29-42 of Algorithm~\ref{alg:alloc}. 
Consider a request to free $blocks$. 
For each unit $u$ containing a \emph{target-to-free} block, mark the free block in $u$ and examine existing states $\textsc{Full}$, $\textsc{Avail}$, and $\textsc{Free}$. 
If $u$ is currently in $\textsc{Full}$, move it to $\textsc{Avail}$ (line 32). 
If there is no allocated block $u$, remove it from $\textsc{Avail}$ and add it to $\textsc{Free}$ (lines 35-36). 
If it is a~\PerfMode{} and there is sufficient memory space in the shared region to accommodate a residual weight (e.g., non-empty \textsc{Evicted}) (line 37), it wakes a state of weight restoration (e.g., \textsc{restoreWeights}).
}

\noindent
\remark{\textbf{Performance Bound on the Switching Heuristic.}}
\dtt{
Our precision-switching scheme can be modeled by a standard discounted Markov Decision Process.
The state is $(k, m)$, where $k \in \{0,...,C_1\}$ is the KV-cache occupancy and $m \in \{0, 1\}$ is the precision mode (0=FP16, 1=FP8). 
The action is whether \textit{stay} or \textit{switch}, and the per-step cost combines latency, accuracy, and a fixed switching cost $\beta$.
Because FP8 enlarges the KV capacity, its advantage in reducing preemption is non-decreasing in $k$. 
This is the monotonicity condition~\cite{puterman2014markov} that implies the optimal value function $V\star$ is submodular in $(k,m)$ and that, with any thresholds $(\kappa_L, \kappa_H)$ that our algorithm picks, the induced policy $\hat\pi$ satisfies the standard policy-search bound~\cite{kakade2002approximately}.
\[
V^{\hat\pi}(s_0) - V^\star(s_0) \;\le\; \frac{\mathcal{E}(\hat\pi)}{1 - \gamma},
\]
where $\mathcal{E}(\hat\pi) \ge 0$ is the per-step \emph{miscalibration cost} of the chosen thresholds---the cost penalty for switching at a slightly wrong KV occupancy---and is zero when the thresholds are placed at the indifference point at which the marginal benefit of switching exactly balances the switching cost~$\beta$.
}

\section{Evaluation}\label{sec:evaluation}

% \begin{table}[t]
% \caption{Workload Characteristics. \todo{Update with our workloads.}}
% \label{tab:workloads}
% \centering
% \resizebox{0.95\columnwidth}{!}{% 표 크기 자동 조절
% \begin{tabular}{@{}lccc@{}}
% \toprule
% Workload & \# Requests & Input Seq. (Mean $\pm$ Std) & Output Seq. (Mean $\pm$ Std) \\ \midrule
% ShareGPT & 2663 & $130.0 \pm 349.9$ & $358.7 \pm 227.4$ \\
% BigCodeBench & 1140 & $172.8 \pm 82.3$ & $155.8 \pm 77.6$ \\
% D1 & 256 & $159.9 \pm 56.5$ & $771.1 \pm 147.0$ \\
% D2 & 512 & $650.3 \pm 220.0$ & $194.1 \pm 37.0$ \\
% D3 & 1024 & $3096.2 \pm 586.1$ & $47.8 \pm 9.5$ \\ \bottomrule
% \end{tabular}%
% }
% % \vspace{-0.1in}
% \end{table}

% \begin{table}[t]
% \small
% \caption{SLO parameters per model.}
% %$T_\text{base}$ is the P99 decode iteration time (batch 32, prefill 4k, no prefill interference).
% %The TBT threshold is $5\times T_\text{base}$; the TTFT threshold is 2\,s.
% %H100 measurements use TP=1 except Phi-3.5-MoE which requires TP=2 (FP16 weights $>$ 80\,GB).}
% \label{tab:slo}
% \centering
% \resizebox{0.7\columnwidth}{!}{%
% \begin{tabular}{@{}lcc@{}}
% \toprule
% & \multicolumn{2}{c}{H100} \\
% \cmidrule(lr){2-3}
% Model & $T_\text{base}$ (ms) & TBT SLO (ms) \\
% \midrule
% Phi-3.5-MoE        & 19.74 & 98.69  \\
% Qwen3-30B-A3B      & 18.19 & 90.93  \\
% GLM-4.7-Flash      & 23.03 & 115.15 \\
% \bottomrule
% \end{tabular}%
% }
% \end{table}

\begin{table*}[t]
    \small
    % \caption{Offline benchmark accuracy across models and serving methods.
    % BF16 is the native model precision.
    % FP8 uses symmetric static per-channel weight quantization with dynamic per-token activation
    % quantization.
    % \sysname \FullMode reconstructs FP16 weights from FP8 base and residual components;
    % \sysname \PerfMode uses only the FP8 base component.
    % The rightmost column reports the mean per-benchmark difference relative to BF16
    % (in percentage points). Per-benchmark variation of $\pm$1--2\,pp is within the noise
    % floor of these benchmarks; small differences should not be interpreted as
    % systematic quality gaps.}
    \caption{Offline benchmark accuracy among BF16, FP8, and \sysname with FP16-only and FP8-only. The rightmost column reports the mean per-benchmark difference relative to BF16 (in percentage points). Per-benchmark variation of $\pm$1--2\,pp is within the noise floor of these benchmarks.}%/; small differences should not be interpreted as systematic quality gaps.}
    \label{tab:accuracy}
    \centering
    \begin{tabular}{@{}ll ccccccc c@{}}
    \toprule
    Model & Method & BBH & GPQA & MATH-500 & MMLU Pro & MUSR & IFEval & LiveCodeBench & \textbf{Avg $\Delta$} \\
    \midrule
    \multirow{4}{*}{Phi-3.5-MoE}
      & BF16      & 74.00 & 36.38 & 37.00 & 60.09 & 45.90 & 64.88 & 22.84 & --- \\
      & FP8       & 76.27 & 35.94 & 38.80 & 59.66 & 45.77 & \textbf{42.70} & 21.80 & $-2.88$ \\
      & \sysname \FullMode  & 73.91 & 35.94 & 36.20 & 60.31 & 46.03 & 66.73 & 23.70 & $+0.25$ \\
      & \sysname \PerfMode  & 74.11 & 35.49 & 37.60 & 59.73 & 47.35 & 65.62 & 22.94 & $+0.25$ \\
    \midrule
    \multirow{4}{*}{Qwen3-30B-A3B}
      & BF16      & 56.40 & 43.53 & 75.80 & 72.42 & 42.46 & 81.70 & 56.21 & --- \\
      & FP8       & 57.32 & 44.87 & 73.40 & 72.54 & 41.27 & 82.07 & 56.02 & $-0.15$ \\
      & \sysname \FullMode  & 58.27 & 41.74 & 75.40 & 72.34 & 41.93 & 81.89 & 55.07 & $-0.27$ \\
      & \sysname \PerfMode  & 58.81 & 42.63 & 74.00 & 72.37 & 42.86 & 82.44 & 57.25 & $+0.26$ \\
    \midrule
    \multirow{4}{*}{GLM-4.7-Flash}
      & BF16      & 81.25 & 37.72 & 20.60 & 63.49 & 41.80 & 80.41 & 34.88 & --- \\
      & FP8       & 80.89 & 33.48 & 22.80 & 62.80 & 42.20 & 79.85 & 34.50 & $-0.52$ \\
      & \sysname \FullMode  & 81.74 & 34.15 & 18.20 & 63.44 & 42.06 & 77.63 & 34.88 & $-1.15$ \\
      & \sysname \PerfMode  & 81.65 & 36.83 & 21.60 & 63.53 & 44.31 & 75.97 & 35.07 & $-0.17$ \\
    \bottomrule
    \end{tabular}%
\end{table*}

\subsection{Methodology} \label{subsec:eval_setup}

\noindent
%\textbf{Experimental Setup.}
We implement \sysname{} on top of vLLM~\cite{kwon2023_vllm} 0.18.0 and evaluate it on three MoE models --- Phi-3.5-MoE~\cite{abdin2024phi3technicalreporthighly}, Qwen3-30B-A3B~\cite{yang2025qwen3technicalreport}, and GLM-4.7-Flash~\cite{5team2025glm45agenticreasoningcoding} --- running on H100 80GB GPUs~\cite{h100}; we use tensor parallelism (TP) when a model exceeds a single device. Specifically, Phi-3.5-MoE uses TP=2.
To emulate bursty serving, we replay two production traces, BurstGPT and Azure: from each we sample a contiguous 2{,}000-request window and scale the arrival rate by $200\times$.
Quality is measured on seven benchmarks --- LiveCodeBench (live coding) plus MMLU-Pro, BBH, GPQA, MATH-500, MUSR, and IFEval (knowledge, reasoning, and instruction following).
% \textbf{Experimental Setup.} 
% \sysname is implemented on top of vLLM~\cite{kwon2023_vllm} 0.18.0.
% We evaluate DPS across three MoE models, two production traces, and seven benchmarks.
% Our model suite includes Phi-3.5-MoE~\cite{abdin2024phi3technicalreporthighly}, Qwen3-30B-A3B~\cite{yang2025qwen3technicalreport}, 
% %DeepSeek-V2-Lite~\cite{deepseekai2024_deepseekv2_strongeconomicalefficient}, Qwen1.5-MoE-A2.7B~\cite{qwen1.5}, 
% and GLM-4.7-Flash~\cite{5team2025glm45agenticreasoningcoding}. 
% To emulate realistic bursty serving conditions, we use two production workload traces, BurstGPT and Azure. For each trace, we sample a contiguous window of 2,000 requests and scale the arrival rate by a factor of 200×. 
% We assess the model's quality across various benchmarks, including LiveCodeBench, and MMLU-Pro, that cover repository-level code understanding, live-coding performance, and general knowledge reasoning, respectively. 
% Experiments are conducted on 
% %NVIDIA A100 40GB~\cite{a100} and 
% H100 80GB GPUs~\cite{h100}. 
% Models that fit within a single device are evaluated in single-GPU mode; for larger models, we use tensor parallelism.

\noindent
\textbf{Baselines.}
We compare \sysname{} against three static-precision configurations --- BF16 (native), static FP8 (symmetric static per-channel weights and symmetric dynamic per-token activations~\cite{vllm-fp8-docs, xiao2023smoothquant, shen2024_efficient_ptq_fp8}), and NestedFP --- plus three baselines that relieve KV pressure through orthogonal mechanisms, all implemented in vLLM~\cite{kwon2023_vllm}.
\kvoff{} evicts least-recently-used KV blocks to pinned CPU memory once the GPU KV pool is exhausted and reloads them on demand.
\kvquant{} stores keys and values in FP8, enlarging the effective KV pool at the cost of dequantization on the read path.
\woff{} keeps a fraction of the weights in pinned CPU memory and streams them in during each forward pass, freeing HBM for additional KV blocks; we use vLLM's offloading mechanism with the per-model configuration that yields the best performance.

\noindent
\textbf{SLO definition.}
Following Sarathi-Serve~\cite{agrawal2024sarathi}, $T_\text{base}$ is the P99 decode iteration time for a batch of 32 requests with 4k context and no prefill interference. 
We measure it on the FP16 configuration, which is \sysname{}'s base mode before any fallback to FP8, so the SLO is consistent across all systems: $T_\text{base} = 19.74$, $18.19$, and $23.03$\,ms on Phi-3.5-MoE, Qwen3-30B-A3B, and GLM-4.7-Flash, respectively. 
A request satisfies the joint SLO iff (1)~its maximum intra-request time-between-tokens (TBT) is $\leq 5\times T_\text{base}$ and (2)~its time-to-first-token (TTFT) is $\leq 2000$~ms. Following DistServe~\cite{zhong2024distserve} and SLOs-Serve~\cite{chen2025slosserve}, system capacity is the highest sustained mean rate at which $\geq 90\%$ of requests satisfy this SLO.

\noindent
\textbf{Metrics.}
Accuracy is task quality on LiveCodeBench and MMLU-Pro; performance is serving behavior under the accelerated BurstGPT~\cite{kdd2025_burstgpt} and Azure traces. 
Because conventional benchmarks score model quality in isolation and ignore latency, we also report \textit{effective pass@1}~\cite{zhong2024distserve, wang2024slo, su2026_morphserve_efficientworkloadawarellm}, the fraction of submitted requests that both satisfy the joint SLO and produce a correct answer per the benchmark's standard pass@1: $\text{effective pass@1} = \text{SLO attainment} \times \text{pass@1}_\text{served}$. 
Requests that miss the SLO contribute zero regardless of accuracy.

% \textbf{Metrics.} 
% \dtt{
% We evaluate DPS from two complementary perspectives: accuracy and performance. On the accuracy side, we compare task quality on  LiveCodeBench, and MMLU-Pro. 
% On the performance side, we measure serving behavior using the accelerated BurstGPT~\cite{kdd2025_burstgpt} and Azure traces, enabling a direct comparison of how DPS trades off quality and efficiency against static FP16 serving, static FP8 serving, and dual-precision NestedFP-style serving. 
% This evaluation design is consistent with prior work that studies the quality-performance tradeoff of adaptive-precision LLM serving under bursty workloads.}
% Conventional benchmarks report model quality in isolation, ignoring whether responses arrive within latency targets. We also report \textit{effective pass@1}~\cite{zhong2024distserve, wang2024slo, su2026_morphserve_efficientworkloadawarellm}: the fraction of submitted requests that both (a) satisfy the joint SLO and (b) produce a correct answer per the standard score of the benchmark (denoted pass@1). 
% Equivalently, $\text{effective pass@1} = \text{SLO attainment} \times \text{pass@1}_\text{served}$.
% This composite metric captures end-to-end user-perceived quality: requests that
% miss the SLO contribute zero, regardless of their accuracy.

\subsection{Experimental Results}

\subsubsection{Accuracy} \label{subsec:accuracy}

% Precision pipeline for each method:
%   BF16       — native model weights, gold-standard quality.
%   FP8        — per-tensor dynamic quantization from BF16.
%   DPS Full   — BF16 → cast FP16 → decompose → reconstruct FP16.
%   DPS Perf   — uses only the FP8 component extracted from the decomposition.
%                Different from FP8 baseline because extraction differs from PTQ.

Table~\ref{tab:accuracy} reports offline benchmark scores for three models under four configurations: BF16 (native precision), static FP8, \sysname{} \FullMode (FP16 reconstructed from the NestedFP decomposition), and \sysname{} \PerfMode (FP8 component only, with non-NestedFP layers retained at original precision). 
The rightmost column gives the mean per-benchmark deviation from BF16 in percentage points (pp).

Both \sysname{} modes preserve BF16 accuracy. 
\FullMode shows mean deltas of $+0.25$, $-0.27$, and $-1.15$\,pp on Phi, Qwen, and GLM, while \PerfMode shows $+0.25$, $+0.26$, and $-0.17$\,pp. 
The \FullMode drift reflects the BF16$\rightarrow$FP16 cast performed at weight loading: NestedFP encodes its base and residual tensors with respect to FP16, so deploying a BF16-native model requires one rounding step that introduces sub-pp drift. 
\PerfMode and \FullMode score within $\pm 0.5$\,pp of one another on average, but this does not mean \PerfMode can replace \FullMode. 
Per-prompt pass@1 is largely insensitive to the precision difference between them, while \FullMode's defining property, producing the same outputs as the deployed FP16 model, is not captured by per-prompt scoring (Section~\ref{subsec:performance} returns to this with online evidence).
Static FP8 shows the largest variance. 
Mean $\Delta$ is modest on Qwen ($-0.15$\,pp) and GLM ($-0.52$\,pp), but Phi-3.5-MoE drops $-2.88$\,pp on average, driven by IFEval where the score falls from $64.88$ to $42.70$. 
Both \sysname{} configurations avoid this collapse on Phi thanks to the partial FP16 retention inherited from NestedFP.

\subsubsection{Performance and Effective Pass@1} \label{subsec:performance}
\begin{figure*}[!t]
  \centering
  \includegraphics[width=0.95\textwidth]{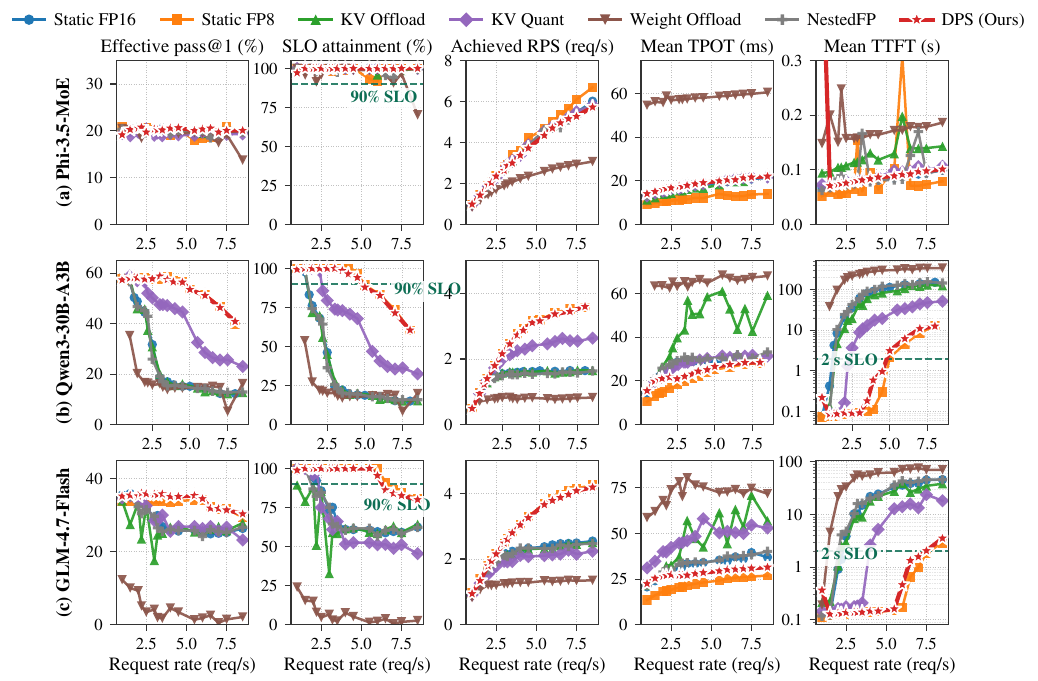}
  \caption{Effective pass@1, SLO attainment, and mean TTFT vs.\ request rate
  on H100 across three MoE models. 
  The 2\,s SLO line is omitted from (a) because all systems remain well below the threshold.}
  \label{fig:main_results}
\end{figure*}

Figure~\ref{fig:main_results} reports the headline result on three models that span low to high KV pressure: Phi-3.5-MoE, Qwen3-30B-A3B, and GLM-4.7-Flash. 
We sweep the request rate from 0.5 to 8.5 req/s under Gamma-distributed arrivals at CV $=2.0$, and we report five metrics: effective pass@1, SLO attainment, achieved RPS, mean TPOT, and mean TTFT. 
We discuss each model separately because the relative behavior of the seven serving methods changes qualitatively with KV pressure.

%\noindent
\textbf{Phi-3.5-MoE: Low-Pressure Regime.}
Phi-3.5-MoE is underloaded over the tested rate range: KV cache never saturates, all seven methods hold $\geq 97\%$ SLO attainment up to 8.5\,req/s, and effective pass@1 sits within a 19.1--21.0\% envelope (2\,pp wide) that reflects benchmark noise rather than method differentiation.
\sysname{} stays inside this envelope at every rate, showing that the SUM operational cost (CUDA VMM mappings, allocator state tracking, controller polling) imposes no measurable accuracy cost when the dynamic mechanism is dormant; 
This addresses the natural concern that \sysname{}'s runtime infrastructure might tax simple workloads.
Throughput tells the same story. Static FP8 leads, \woff{} trails, and \sysname{}, static FP16, and NestedFP stay within 15\% of static FP8. In this regime, \sysname{} reduces to its base configuration: FP16 execution with no mode transitions, and it matches static FP16 on every operational metric.
This also serves as a multi-GPU sanity check, since Phi-3.5-MoE requires TP$=2$ on H100: \sysname{}'s per-rank shared regions and SUM Manager broadcasts add nothing on top of the tensor-parallel cost that static FP16 already pays.

% \textbf{Phi-3.5-MoE: Low-Pressure Regime}.
% Phi-3.5-MoE represents the underloaded regime where the KV cache never saturates within the tested rate range. 
% All seven methods sustain $\ge$97\% SLO attainment up to 8.5 req/s, and effective pass@1 stays within 19.1–21.0\% across the entire sweep — a 2pp envelope that reflects benchmark noise rather than method differentiation. 
% \sysname sits within this envelope at every rate, demonstrating that the SUM machinery (CUDA VMM mappings, allocator state tracking, controller polling) introduces no measurable accuracy regression when the dynamic mechanism is not engaged. 
% This negative-control result preempts the natural concern that DPS's runtime infrastructure might tax simple workloads.
% The throughput result is rather identical to the accuracy result. 
% Since the pressure is very low, every method sustains the full sweep of request rates while keeping SLO attainment around 99\%. 
% Static FP8 achieves the highest throughput, \woff performs the worst, and all other methods behave similarly. The throughputs of \sysname, static FP8, static FP16, and NestedFP are within 15\% of one another. 
% In this regime, \sysname effectively reduces to its base configuration — full-quality (FP16) execution with no mode transitions — and the system tracks Static FP16 closely on every operational metric.

%\noindent
\textbf{Qwen3-30B-A3B: High-Pressure Regime.}
Qwen3-30B-A3B is the regime where Static FP16 fails most dramatically and \sysname{}'s dynamic mechanism contributes most. 
At 1.0\,req/s (below saturation), \sysname{}, Static FP16, Static FP8, and NestedFP all deliver 56.7--58.0\% effective pass@1, confirming that \sysname{} preserves FP16-class quality at low load. 
As the rate increases, the curves diverge sharply: by 2.5\,req/s, Static FP16 has collapsed to 31.8\% effective pass@1 (from 56.8\%), driven by SLO attainment falling to 46.2\%, while \sysname{} sustains 57.8\% with full SLO compliance. By 4.5\,req/s, \sysname{} delivers 56.4\% against Static FP16's 15.2\%, a $+41.2$\,pp absolute advantage.

The Static FP16 collapse is not a quality failure: its unconditional pass@1 at 4.5\,req/s remains $\sim 57.4\%$. 
The collapse stems from SLO violations: Static FP16 has limited KV cache capacity, and under pressure, many requests violate the SLO, contributing zero to effective pass@1. 
NestedFP, \kvoff{}, and \kvquant{} collapse similarly; \kvquant{} degrades more gracefully because per-token KV reduction expands effective capacity, but still trails \sysname{} by 9--14\,pp across the burst region. 
Static FP8 tracks \sysname{} throughout (56.5\% vs.\ 56.4\% at 4.5\,req/s), but pays its quality cost permanently rather than only during bursts.
Throughput and latency expose the mechanism. 
Static FP16 sustains only 0.95 RPS at 1.0\,req/s before SLO collapses, whereas \sysname{} sustains 3.11 RPS at 4.5\,req/s, a $3.3\times$ extension of FP16's viable range that matches Static FP8 (3.16 RPS) without paying FP8's accuracy cost. 
NestedFP gains nothing over Static FP16 (0.94 RPS) because its GPU memory is used to store the dual-precision weight. 
TTFT tells the most dramatic story: Static FP16 climbs from 428\,ms at 1.0\,req/s to 8.4\,s at 1.5 and 21.0\,s at 2.0, while \sysname{} stays at 88--93\,ms up to 3.0\,req/s (vs.\ 61\,s for Static FP16, a $\sim$660$\times$ reduction) and remains within the 2\,s prefill threshold (0.97\,s) even at its terminal SLO-feasible rate of 4.5\,req/s. 
TPOT confirms a small per-token reconstruction cost: Static FP8 is fastest (13--24\,ms) on its smaller weight footprint, while \sysname{} in full-quality mode tracks Static FP16 within 1--3\,ms (20--28 vs.\ 19--30\,ms), bounded regardless of load.

\begin{figure*}[t]
      \centering
     \includegraphics[width=0.80\linewidth]{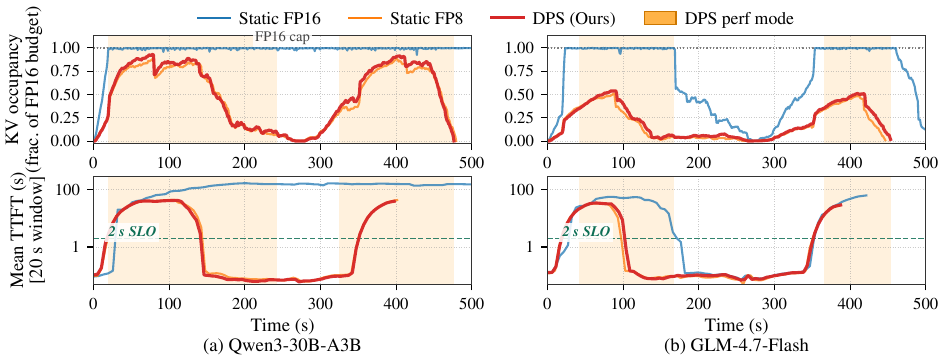}
     \caption{KV cache occupancy (top) and mean TTFT (bottom) of \sysname on (a) Qwen3-30B-A3B and (b) GLM-4.7-Flash.}
      %\caption{\sysname on (a) Qwen3-30B-A3B and (b) GLM-4.7-Flash. Top: KV cache occupancy (fraction of FP16 budget). Bottom: mean TTFT (20\,s rolling window).}
      \label{fig:dps_memory_elasticity}
\end{figure*}

%\noindent
\textbf{GLM-4.7-Flash: Moderate-Pressure Regime.}
GLM-4.7-Flash repeats Qwen's pattern shifted right on the rate axis, since its smaller per-request KV footprint slows the buildup of KV pressure. 
At 1.0\,req/s, all methods are near-equivalent (\sysname{} 35.3\%, Static FP16 35.6\%, Static FP8 33.9\%). 
Static FP16, NestedFP, and \kvoff{} start failing the SLO around 2.5\,req/s, and their effective pass@1 falls from 35\% to 25\% by 4.5\,req/s purely from SLO violation (unconditional pass@1 stays near 35\%). 
\sysname{} sustains 35.2\% at 4.5\,req/s with full SLO compliance: a $+9.5$\,pp gain over Static FP16, and $+1.3$\,pp over Static FP8 (33.9\%) because \sysname{} spends part of the low-load intervals in full-quality mode.
Throughput follows the same pattern: Static FP16 caps at 1.80 RPS at 2.2\,req/s while \sysname{} sustains 3.78 RPS at 6.0\,req/s, a $2.1\times$ extension. 
NestedFP again matches Static FP16 (1.34 RPS) for the same structural reason as on Qwen.

On the latency side, mean TTFT for static FP16 climbs from 910 ms at req/s of 2.0 to 10.3 s at 3.0, while \sysname holds at 130–147 ms up to req/s of 4.5 (vs. 24 s for static FP16, a roughly 165 $\times$ reduction). \sysname's mean TTFT remains within the 2-second prefill threshold throughout its operating range (0.46 s at req/s of 6.0), while FP16 has long since left it.
TPOT splits the methods into three tiers: Static FP8 is fastest (14--26\,ms), 
\sysname{} sits slightly higher due to its reconstruction cost (22--30\,ms), 
and Static FP16 is slowest (21--35\,ms). 
The offloading baselines degrade much more sharply under load: \kvoff{} climbs to 71\,ms because every decode step fetches KV blocks from CPU memory, and \woff{} stays permanently above 60\,ms because every forward pass touches a fraction of host-resident weights. \sysname{}'s background restore stays off the decode hot path: residual transfers run on a separate copy stream and mode transitions are batch-aligned (Section~\ref{subsec:KV-aware}).

% \begin{table}[t]
% \small
%   \caption{Mode switching overhead under bursty traces measured on a BurstGPT window.
%   Restore latency is the per-cycle mean of the weight H2D wall time.}
%   \label{tab:switching_overhead}
%   \centering
%   \resizebox{0.95\columnwidth}{!}{%
%   \begin{tabular}{@{}l cc cc c@{}}
%   \toprule
%    & \multicolumn{2}{c}{Time in mode (\%)} & \multicolumn{2}{c}{Transitions} & Restore \\
%   \cmidrule(lr){2-3} \cmidrule(lr){4-5}
%   Model & Full & Perf & F$\to$P & P$\to$F & Latency (ms) \\
%   \midrule
%   Phi-3.5-MoE        & 100.0 & 0.0   & 0  & 0  & --     \\
%   Qwen3-30B-A3B      & 20.8  & 79.2  & 2  & 2  & 24651  \\
%   %DeepSeek-V2-Lite   & --    & --    & -- & -- & --     \\
%   %Qwen1.5-MoE-A2.7B  & --    & --    & -- & -- & --     \\
%   GLM-4.7-Flash      & 52.9  & 47.1  & 2  & 2  & 8497   \\
%   \bottomrule
%   \end{tabular}%
%   }
% \end{table}

%\noindent
%\textbf{Contribution of \FullMode.}
\textbf{Impact of \FullMode.}
Table~\ref{tab:accuracy} shows \FullMode and \PerfMode within noise offline, raising the question of \FullMode's contribution. 
To isolate per-request output quality, we compare \sysname{} against static FP8 only where both clear $\geq 95\%$ SLO attainment, so neither is dropping requests. 
In this regime, \sysname{}'s unconditional pass@1 exceeds static FP8's by $+1.39$\,pp on GLM, $+0.46$\,pp on Qwen, and $+0.32$\,pp on Phi, winning on $30$ of $35$ (model, req/s) combinations.
Additionally, on Qwen, \sysname{}'s online unconditional pass@1 ($\sim$58\%) closely matches \PerfMode's offline LiveCodeBench score ($57.25\%$), so the win over static FP8 is explained by \PerfMode's FP16 retention alone.
On GLM, \sysname{}'s online score ($35.2$--$35.7\%$) clearly exceeds \PerfMode's offline score ($35.07\%$), and the online gap over static FP8 ($+1.39$\,pp) is substantially larger than the offline \PerfMode-vs-FP8 gap ($+0.57$\,pp).
Therefore, the residual $\sim$0.8--$1.0$\,pp on GLM reflects the contribution of \FullMode in the online scenario.
The rate trend confirms it: the gap is largest at low rates ($+1.83$\,pp at $\lambda = 1.0$) where \sysname spends most time in \FullMode, and shrinks to zero at high rates ($-0.03$\,pp at $\lambda = 7.5$) as KV pressure forces the system to switch to ~\PerfMode.
%\FullMode's empirical contribution is therefore model-dependent, but its principled value --- producing FP16-equivalent outputs for reproducibility and audit --- applies regardless.

% \begin{figure*}[t]
%       \centering
%      \includegraphics[width=1.0\linewidth]{figure/dps_memory_elasticity.pdf}
%      \caption{KV cache occupancy (top) and mean TTFT (bottom) of \sysname on (a) Qwen3-30B-A3B and (b) GLM-4.7-Flash.}
%       %\caption{\sysname on (a) Qwen3-30B-A3B and (b) GLM-4.7-Flash. Top: KV cache occupancy (fraction of FP16 budget). Bottom: mean TTFT (20\,s rolling window).}
%       \label{fig:dps_memory_elasticity}
% \end{figure*}

\noindent
\textbf{Summary}.
Across the three regimes, a consistent story emerges.
At low load, \sysname{} pays a few-millisecond TPOT overhead and at most 25\,ms TTFT overhead for SUM bookkeeping and FP16 reconstruction. 
Once KV pressure binds, \sysname{} delivers 2.1--3.3$\times$ higher sustained throughput, two-orders-of-magnitude TTFT improvements over Static FP16, and better effective pass@1. 
Under KV pressure, \sysname{} 's elastic KV capacity reduces preemption, lets \sysname{} keep larger active batches, and amortizes per-step weight reads, while Static FP16's fixed budget cannot hold a full batch. 
Static FP8 is the only baseline matching \sysname{} on throughput and SLO compliance, but \sysname{} still gains 1--3\,pp effective pass@1 on GLM (from time spent in full-quality mode at low and moderate rates) at a small TPOT cost. 
No single static configuration wins across regimes: Static FP8 wins TPOT at the cost of permanent accuracy loss, Static FP16 wins TPOT at low load but collapses under bursts, and \sysname{} wins on effective pass@1 across the entire range.

% \begin{table}[t]
% \small
%   \caption{Mode switching overhead under bursty traces measured on a BurstGPT window.
%   Restore latency is the per-cycle mean of the weight H2D wall time.}
%   \label{tab:switching_overhead}
%   \centering
%   \resizebox{0.95\columnwidth}{!}{%
%   \begin{tabular}{@{}l cc cc c@{}}
%   \toprule
%    & \multicolumn{2}{c}{Time in mode (\%)} & \multicolumn{2}{c}{Transitions} & Restore \\
%   \cmidrule(lr){2-3} \cmidrule(lr){4-5}
%   Model & Full & Perf & F$\to$P & P$\to$F & Latency (ms) \\
%   \midrule
%   Phi-3.5-MoE        & 100.0 & 0.0   & 0  & 0  & --     \\
%   Qwen3-30B-A3B      & 20.8  & 79.2  & 2  & 2  & 24651  \\
%   %DeepSeek-V2-Lite   & --    & --    & -- & -- & --     \\
%   %Qwen1.5-MoE-A2.7B  & --    & --    & -- & -- & --     \\
%   GLM-4.7-Flash      & 52.9  & 47.1  & 2  & 2  & 8497   \\
%   \bottomrule
%   \end{tabular}%
%   }
% \end{table}

\subsubsection{DPS Characterization} \label{subsec:memory}
This section traces the internal state behind those outcomes during a representative window, run at $4.5$\,req/s on Qwen3-30B-A3B and $6.0$\,req/s on GLM-4.7-Flash --- rates where Static FP16 has collapsed but \sysname sustains the SLO.
Figure~\ref{fig:dps_memory_elasticity} reports two signals on a shared time axis: KV cache occupancy (top) and mean TTFT (bottom).

% \begin{figure}[t]
%       \centering
%      \includegraphics[width=1.0\linewidth]{figure/dps_memory_elasticity.pdf}
%       \caption{\sysname on (a) Qwen3-30B-A3B and (b) GLM-4.7-Flash. Top: KV cache occupancy (fraction of FP16 budget). Bottom: mean TTFT (20\,s rolling window).}
%       \label{fig:dps_memory_elasticity}
% \end{figure}

\noindent
\textbf{KV occupancy}.
Static FP16 stays at the FP16 budget for nearly the whole window.
Once the cap is reached, every burst arrival must queue or preempt.
\sysname avoids this by reclaiming residual-weight units: KV occupancy oscillates between $0$ and $0.93$ on Qwen and $0$ and $0.54$ on GLM, with peaks aligned to bursts.
The amber bands show \sysname enters performance mode exactly when KV demand exceeds the FP16 budget --- about 79.2\% of the trace on Qwen, 47.1\% on GLM --- and reverts to full-quality mode when pressure subsides.
\sysname evicts up to 40 of 40 residual units on Qwen (the entire shared region) and 14 of 40 on GLM, reflecting GLM's smaller per-request KV footprint.

\noindent
\textbf{TTFT}.
Static FP16's mean TTFT rises up to 170\,s on Qwen and 60\,s on GLM --- two orders of magnitude above the 2\,s SLO, which is caused by the KV pressure.
In two models, \sysname crosses the SLO only at the trace start and end; during steady serving, TTFT remains below the SLO. %The amber bands align with TTFT recovery: when \sysname{} enters performance mode and reclaims memory, queue depth drops and TTFT returns to interactive levels.
Static allocation turns a transient memory spike into a long throughput collapse, since preemption cannot enlarge the budget.
\sysname turns the same spike into a brief precision excursion, with capacity restored once the burst passes.
The $2.7$--$4.5\times$ operating-range extension and up to $+41$\,pp effective pass@1 reported in Section~\ref{subsec:performance} follow directly from this observation.

\subsubsection{Mode Switching Overhead} \label{subsec:switching}

Table~\ref{tab:switching_overhead} summarizes mode-switching activity under the bursty traces of Section~\ref{subsec:performance}.
\sysname{} spends 100\% of serving time in full-quality mode on Phi-3.5-MoE (no KV pressure, no transitions), matching the no-overhead behavior in Figure~\ref{fig:main_results}(a).
On Qwen3-30B-A3B and GLM-4.7-Flash, only two full~$\to$~perf and two perf~$\to$~full transitions occur across the entire trace, showing that the threshold-with-hysteresis policy (Section~\ref{subsec:KV-aware}) consolidates bursts into a few long mode-windows rather than flapping rapidly.
The reverse transition dominates the cost: average end-to-end restore latency is 8.5\,s on GLM and 24.7\,s on Qwen, set by the H2D transfer of 14/40 and 40/40 residual units from pinned CPU memory at near-identical per-unit cost (607 vs.\ 616\,ms/unit) --- restoration thus scales linearly with evicted units rather than total model size, and GLM's lower total reflects only its lighter peak pressure.
Because this transfer runs on the background restore worker and overlaps with ongoing decode, it does not block requests; its only runtime cost is bounded PCIe contention, and the user-visible impact remains negligible (Section~\ref{subsec:performance}).

\begin{table}[t]
\small
  \caption{Mode switching overhead under BurstGPT window's traces.
  Restore latency is the per-cycle mean of the weight H2D wall time.}
  \label{tab:switching_overhead}
  \centering
  \resizebox{0.95\columnwidth}{!}{%
  \begin{tabular}{@{}l cc cc c@{}}
  \toprule
   & \multicolumn{2}{c}{Time in mode (\%)} & \multicolumn{2}{c}{Transitions} & Restore \\
  \cmidrule(lr){2-3} \cmidrule(lr){4-5}
  Model & Full & Perf & F$\to$P & P$\to$F & Latency (ms) \\
  \midrule
  Phi-3.5-MoE        & 100.0 & 0.0   & 0  & 0  & --     \\
  Qwen3-30B-A3B      & 20.8  & 79.2  & 2  & 2  & 24651  \\
  %DeepSeek-V2-Lite   & --    & --    & -- & -- & --     \\
  %Qwen1.5-MoE-A2.7B  & --    & --    & -- & -- & --     \\
  GLM-4.7-Flash      & 52.9  & 47.1  & 2  & 2  & 8497   \\
  \bottomrule
  \end{tabular}%
  }
\end{table}

\section{Related Works and Discussion}\label{sec:related_work}
%% NXT ++ 20260408
% \noindent
% \textbf{Dynamic Neural Networks.}
% \sysname{} specifically targets \emph{dynamic neural networks}, including nested FP~\cite{lee2026_nestedfp}, any precision networks~\cite{yu_aaai_2021_dynamic, icml24_anyprecision}, early-exit networks~\cite{Jeon_2024_WACV_ee, chen_icml_2024_ee, xu_isca_2025_ee}. 
% These networks can typically adapt their structures or parameters and select an execution path based on a specific condition, for example, sample-wise difficulty~\cite{yu_aaai_2021_dynamic, Jeon_2024_WACV_ee} or a computational requirement~\cite{lee2026_nestedfp, icml24_anyprecision}. 
% Built on top of dynamic networks,~\emph{Semi-Unified Memory} partitions the weight region into a persistent sub-region and a shared region that alternates between residual weight tensors and KV-cache blocks, effectively leveraging both a~\PerfMode{} and a~\FullMode{}.

\noindent
\textbf{LLM Serving Systems with Virtual Memory.}
One fundamental feature of~\sysname{} is to preserve paged KV cache management in modern LLM-serving systems, such as in vLLM~\cite{kwon2023_vllm, kwon2023_pageattention}. 
%In particular, inspired by classical virtual memory and paging techniques in operating systems, vLLM facilitates flexible sharing of the KV cache within and across requests, effectively reducing memory usage and thereby achieving near-zero waste in KV cache memory. 
Some recent approaches, such as~\cite {yu2025_prism, su2026_morphserve_efficientworkloadawarellm}, also leverage dynamic memory management when serving multiple LLM models. 
~\cite{yu2025_prism} primarily targets memory sharing across different vLLM engines, while~\cite{su2026_morphserve_efficientworkloadawarellm} focuses on serving with a quantized model and an FP16 model. 
%Meanwhile, built on \emph{semi-unified memory}, ~\sysname{} preserves the powerful KV management of existing serving systems while providing flexible runtime memory allocation.  

\noindent
\textbf{Model Offloading.} %~\sysname{} also requires offloading a partial model from CPU memory to GPU memory. 
~\sysname{} performs offloading of partial model state from CPU to GPU during full-quality-mode restoration.
There are many recent offloading techniques, such as those in~\cite{sheng_icml23_flexgen, cao_asplos2025_moe_lightning, su2026_morphserve_efficientworkloadawarellm}, for addressing GPU memory constraints in LLM serving systems. 
A common concept is to hide host-to-device memory transfers during GPU computation. 
For example,~\cite{sheng_icml23_flexgen} executes a layer in batches, providing sufficient GPU computation time to hide model transfers. 
~\cite{su2026_morphserve_efficientworkloadawarellm} assumes that model transfers can be hidden by GPU computation. 
%In~\sysname{}, offloading a partial model is used when restoring a~\FullMode{} from a~\PerfMode{}.
%As a result, a performance mode is still maintained until the completion of offloading a partial model, ensuring a smooth mode transition.

\noindent
\textbf{Limitations and Future Work.} Page-based KV memory management may incur memory fragmentation~\cite{prabhu_vAttention_asplos_2025, xu2025_ellm_elastic_memorymanagement}. 
Meanwhile, the nested FP8-FP16 model is a specific case of dynamic networks. 
%Inspired by vLLM~\cite{kwon2023_vllm, kwon2023_pageattention},~\sysname{} primarily aims to enable flexible sharing of the KV cache within and across requests, thereby increasing the utilization of KV cache blocks. 
%However, paged KV memory management may incur memory fragmentation~\cite{prabhu_vAttention_asplos_2025, xu2025_ellm_elastic_memorymanagement}. 
%Meanwhile,~\sysname{} specifically presents dual-moserving as an example of SUM. 
Investigating KV cache fragmentation or extending the method to various dynamic networks are promising future directions for~\sysname{} with SUM. 
Additionally,~\sysname{} focuses mainly on a coarse-grained nested model, as our target is to immediately expand a page-based KV cache as in vLLM. 
This design can be extended to layer-wise weight tensors, as in some recent layer-wise mixed-precision approaches~\cite{su2026_morphserve_efficientworkloadawarellm, mai_udp}, which can greatly extend the flexibility of~\sysname{}.
We treat this integration as promising future work. 

\section{Conclusion}\label{sec:conclusion}
\remark{This paper introduces an effective dual-precision serving with semi-unified memory. 
The proposed scheme is built on two fundamental concepts: (1) SUM to effectively mitigate KV cache pressure and possibly reduce preemptions, and (2) dual-precision serving with asymmetric switching based on nested models. 
With SUM, it turns weight memory into an elastic resource: under normal load, DPS serves the higher-precision model; under KV pressure, it switches to a nested, lower-precision variant and repurposes unused weight memory for KV cache blocks. 
\sysname{} can serve as an extension for existing LLM serving systems, such as vLLM~\cite{kwon2023_pageattention} while maintaining the effective dynamic structure of KV cache management. 
}

\bibliographystyle{IEEEtran}
\bibliography{references}

\vspace{12pt}

% \color{red}
% IEEE conference templates contain guidance text for composing and formatting conference papers. Please ensure that all template text is removed from your conference paper prior to submission to the conference. Failure to remove the template text from your paper may result in your paper not being published.

\end{document}